\documentclass[12pt,preprint]{aastex}

\usepackage{natbib}
\usepackage[utf8]{inputenc}
\usepackage{graphicx}
\usepackage{amsmath}
\usepackage{lscape}

\begin{document}

\title{Nearby Young, Active, Late-type Dwarfs in {\it Gaia}'s First Data Release} 

\author{Joel H.\ Kastner\altaffilmark{1}, Germano Sacco\altaffilmark{2}, David Rodriguez\altaffilmark{3}, Kristina Punzi\altaffilmark{1}, B. Zuckerman\altaffilmark{4}, Laura Vican Haney\altaffilmark{4}}

\altaffiltext{1}{Chester F. Carlson Center for Imaging Science, School of Physics \& Astronomy, and 
Laboratory for Multiwavelength Astrophysics, Rochester Institute of
Technology, 54 Lomb Memorial Drive, Rochester NY 14623 USA
(jhk@cis.rit.edu)}
\altaffiltext{2}{INAF-Osservatorio Astronomico di Arcetri, Largo E. Fermi, 5, 50125, Firenze, Italy }
\altaffiltext{3}{Department of Astrophysics, American Museum of Natural History, Central Park West at 79th Street, New York, NY 10034, USA}
\altaffiltext{4}{Department of Physics \& Astronomy, UCLA, Los Angeles, CA}

\begin{abstract}
The {\it Galex} Nearby Young Star Survey (GALNYSS) has yielded a sample of $\sim$2000 UV-selected objects that are candidate nearby ($D \stackrel{<}{\sim}$150 pc), young (age $\sim$10--100 Myr), late-type stars. 
Here, we evaluate the distances and ages of the subsample of (19) GALNYSS stars with {\it Gaia} Data Release 1 (DR1) parallax distances $D \le 120$ pc. The overall youth of these 19 mid-K to early-M stars is readily apparent from their positions relative to the loci of main sequence stars and giants in {\it Gaia}-based color-magnitude and color-color diagrams constructed for all stars detected by {\it Galex} and the {\it Wide-field Infrared Space Explorer} for which parallax measurements are included in DR1.
The isochronal ages of all 19 stars lie in the range $\sim$10--100 Myr. Comparison with Li-based age estimates indicates a handful of these stars may be young main-sequence binaries rather than pre-main sequence stars. Nine of the 19 objects have not previously been considered as nearby, young stars, and all but one of these are found at declinations north of $+$30$^\circ$. 
The {\it Gaia} DR1 results presented here indicate that the GALNYSS sample includes several hundred nearby, young stars, a substantial fraction of which have not been previously recognized as having ages $\stackrel{<}{\sim}$100 Myr.
\end{abstract}

\section{Introduction}

The past two decades have seen dramatic progress in our knowledge of the population of stars of age $\stackrel{<}{\sim}$200 Myr that lie within $\sim$100 pc of the Sun \citep[][]{ZuckermanSong2004,Torres2008,Kastner2016IAUSed}. Such nearby, young stars, most of which are found in loose kinematic groups \citep[for a recent review, see][]{Mamajek2016}, provide unique tests of pre-main sequence stellar evolution and the late stages of evolution of protoplanetary disks  \citep[][and refs.\ therein]{Kastner2016IAUScont}, and represent the best targets for direct imaging searches for giant exoplanets \citep[e.g.,][]{Chauvin2016}. 

The majority of stars of type F through early M that are members of these nearby young moving groups (NYMGs) have been identified, or their youth has been confirmed, via their anomalously large coronal X-ray fluxes relative to the late-type field star population \citep[][and refs.\ therein]{ZuckermanSong2004,Torres2008}. This intense coronal emission is the external manifestation of the strong internal magnetic dynamos and, hence, strong surface magnetic fields that  are characteristic of young, late-type stars. If the mass functions of NYMGs resemble those of the field, however, then the
presently known membership of these nearby groups is significantly
lacking in stars of spectral type later than K \citep[e.g.,][]{Gagne2017}. This deficiency is due, in part, to a
lack of sensitive X-ray measurements over
sufficiently large regions of sky, such as would be necessary to detect nearby young stars with
spectral types later than early M at distances beyond a few tens of
pc \citep{Rodriguez2013}. 

Fortunately, the same strong stellar surface magnetic activity that generates X-ray-luminous coronae around young stars also leads to UV-luminous chromospheres. Hence, young late-type stars also stand out against field dwarfs in terms of their UV emission.
The {\it Galex}  (UV) All-sky Imaging Survey \citep[AIS;][]{Bianchi2014}, which covered $\sim$90\% of the sky in its far-ultraviolet (FUV, 1344--1786 \AA) and near-ultraviolet (NUV, 1771--2831 \AA) bandpasses, thereby represents a powerful means to identify young K and M stars in the field \citep[][]{Shkolnik2011}. The {\it Galex} AIS data are rendered particularly powerful when coupled with proper motion and/or radial velocity (RV) data that can constrain the space motions (i.e., Galactic $U,V,W$ velocity components) of candidate stars \citep[][]{Rodriguez2011}.

To exploit this potential, we \citep[][]{Rodriguez2011,Rodriguez2013} developed a method to select candidate nearby, young, late-type stars on the basis of their UV excesses and NYMG-like space motions. This method involves analysis of the results of the cross-correlation of the  {\it Galex} AIS source catalog with proper motion catalogs and the all-sky infrared point source catalogs compiled from Two Micron All-sky Survey (2MASS) and {\it Wide-field Infrared Space Explorer} (WISE) data. The resulting {\it Galex} Nearby Young Star Survey (GALNYSS) sample consists of $\sim$2000 candidate young (age $\sim$10--100 Myr), nearby (D$\stackrel{<}{\sim}$150 pc), late-type stars \citep{Rodriguez2013}. 

As a consequence of the faintness of field K and M dwarfs at distances beyond a few tens of pc, the vast majority of the GALNYSS stars lack parallax measurements. Hence, \citet[][]{Rodriguez2013} estimated the likely distance range of each star based on the assumption that its age lies in the range 10--100 Myr. Precise distance determinations are required to confirm (or refute) the young-star status of individual GALNYSS sample objects and, for those stars confirmed as young, to establish their likely ages. Furthermore, the available proper motion data is of relatively low quality for a significant fraction of these stars; in such cases, we calculated proper motions from 2MASS and WISE positions, yielding proper motions that are typically quite uncertain.  Thus --- while preliminary investigations suggest that a significant fraction of GALNYSS sample stars are indeed young and nearby \citep{Rodriguez2011,Rodriguez2013} --- the precise distances and ages of most of the individual stars, and their status as members (or nonmembers) of known NYMGs, remain to be determined.

The forthcoming avalanche of high-precision parallaxes and proper motions for $\sim$$10^9$ stars that will flow from the {\it Gaia} mission \citep{Gaia2016} will enable confirmation or refutation of nearby, young star status for the vast majority of the $\sim$2000 GALNYSS candidates identified thus far. 
{\it Gaia} Data Release (DR1), which includes the {\it Tycho} {\it Gaia} Astrometric Solution (TGAS) data --- consisting of parallaxes and proper motions for $\sim$2.5$\times10^6$ stars \citep{GaiaDR12016} --- provides the first opportunity to use {\it Gaia} data to test the youth and proximity of GALNYSS stars. Here, we present a ``proof of concept'' study that  makes use of the {\it Gaia} DR1 data to verify the young-star status, and infer ages and space motions, for the small fraction ($\sim$2.5\%) of the GALNYSS sample that is included in TGAS.


\section{Sample: GALNYSS Stars with {\it Gaia}-based Parallax Distances $\le$120 pc}

We used the Vizier cross-match tool\footnote{http://cdsxmatch.u-strasbg.fr/xmatch} to identify all stars in the GALNYSS sample of 2031 objects compiled by \citet{Rodriguez2013} with parallax ($\pi$) measurements in the {\it Gaia} TGAS catalog. This cross-match exercise, which used a matching radius of 3$''$, generated a total of 50 objects. The small fraction of GALNYSS objects with parallax and proper motion data in TGAS (50 of $\sim$2000, i.e., $\sim$2.5\%) most likely reflects the overall faintness of the GALNYSS sample, which has a mean $V$ mag $\sim$14 (compared with a mean $G$ mag of $\sim$10.8 for TGAS). 

To further limit the initial sample studied here to those GALNYSS stars included in TGAS that are closest to Earth, we then selected only those objects for which $\pi\ge8$ mas, corresponding to distances $D\le120$ pc --- i.e., closer than the nearest known actively star-forming molecular cloud complexes (such as those in Taurus-Aurigae and Ophiuchus). 
This straightforward selection criterion ($\pi\ge8$ mas) yielded the subsample of 19 GALNYSS objects whose  {\it Gaia} TGAS catalog data (positions, parallaxes, proper motions, and $G$ magnitudes) are listed in Table~\ref{tbl:GaiaData}. For the remaining 31 GALNYSS objects present in TGAS, the {\it Gaia} parallaxes are either unreliable (e.g., have high relative uncertainties) or are too small to be consistent with the hypothesis that these objects are stars (young or otherwise) within $\sim$120 pc. 

\subsection{Distances and bolometric luminosities}

In Table~\ref{tbl:DerivedData}, we list the stellar distances corresponding to the parallaxes in Table~\ref{tbl:GaiaData}, as well as the $G$ magnitudes and {\it Galex/Gaia/2MASS} colors of the Table~\ref{tbl:GaiaData} stars. We used the distances and the listed spectral types to estimate bolometric luminosities $L_{\rm bol}$ from the stars' $V$ magnitudes \citep[as obtained from {\it Gaia}  pre-launch, spectral type-dependent $G$ vs.\ $V$ relations in][]{Jordi2010} and the spectral type-dependent bolometric corrections tabulated in \citet{PecautMamajek2013}. For the handful of stars for which no spectral type is available in the literature, the spectral types, and hence $L_{\rm bol}$ values, are based on consideration of $G-K_S$ colors (see \S\ 3.2) as well as {\it 2MASS} \citep{Skrutskie2006} and {\it WISE} \citep{Wright2010} colors, following \citet{PecautMamajek2013}. 

\subsection{Relative X-ray luminosities}

To further investigate the pre-MS status of the Table~\ref{tbl:GaiaData} stars, we consulted the HEASARC archives\footnote{https://heasarc.gsfc.nasa.gov/db-perl/W3Browse/w3browse.pl} for each star's association (or lack thereof) with an X-ray source in the {\it ROSAT} All-Sky Survey (RASS). We assume a RASS source is associated with a given star if the source-star angular offset is $\le$30$''$, i.e., smaller than the nominal PSF of {\it ROSAT}'s Position Sensitive Proportional Counter. We find the majority of the Table~\ref{tbl:GaiaData} stars have associated RASS X-ray sources. Only one star (HIP 59748) is a RASS nondetection, and there are five cases of ambiguous RASS source associations, i.e., wide binaries or low-significance RASS sources lying at offsets $>$30$''$ from Table~\ref{tbl:GaiaData} stars (see Table~\ref{tbl:DerivedData} footnotes). 

The count rates of RASS sources unambiguously associated with Table~\ref{tbl:GaiaData} stars were converted to fluxes assuming a conversion factor appropriate for young late-type stars, i.e., $6.3\times10^{-12}$ erg cm$^{-2}$ count$^{-1}$ \citep{Kastner2016}. The resulting X-ray fluxes were used to determine the stars' X-ray luminosities relative to bolometric, $\log{(L_X/L_{\rm bol})}$, at  the epoch of the RASS observations. These $\log{(L_X/L_{\rm bol})}$ values are listed in Table~\ref{tbl:DerivedData}. For all but one of the Table~\ref{tbl:GaiaData} stars with unambiguous RASS detections or nondetections, $\log{(L_X/L_{\rm bol})}$ lies in the range $-4 \stackrel{<}{\sim} \log{(L_X/L_{\rm bol})} \stackrel{<}{\sim} -3$, as expected for late-K and early-M pre-MS stars \citep[see, e.g.,][and refs.\ therein]{Kastner2016}. The lone exception is HIP 59748, whose RASS upper limit suggests $\log{(L_X/L_{\rm bol})} < -4.4$ (see \S\ \ref{sec:Imposters}).

\subsection{Spectroscopy}

The GALNYSS sample stars have been the subject of an ongoing optical/IR spectroscopic survey \citep[][]{Rodriguez2013,Vican2016}. To date, spectra have been obtained for 13 of the 19 Table~\ref{tbl:GaiaData} stars. We list the telescopes, instruments, and observing dates for these observations, as well as results for radial velocities (RVs) and equivalent widths (EWs) of H$\alpha$ emission and Li {\sc i} $\lambda$6708 absorption lines, in Table~\ref{tbl:Obs}.
We make use of these RV and EW measurements to calculate space ($UVW$) velocities and in assessing stellar age and NYMG membership, as summarized in Table~\ref{tbl:Summary} and described in \S\S\ \ref{sec:XYZUVW}, \ref{sec:Individuals}. Details concerning spectrum acquisition, reduction, and analysis will be presented in a forthcoming paper \citep{VicanHaney2017}. 

\subsection{Galactic positions and space velocities \label{sec:XYZUVW}}

The availability of precise parallaxes in the {\it Gaia} TGAS catalog yields improved Galactic positions ($X, Y, Z$) for the Table~\ref{tbl:GaiaData} stars, while the availability of {\it Gaia} TGAS proper motions provides the opportunity to recalculate $U, V, W$ space velocities for the 18 Table~\ref{tbl:GaiaData} stars whose RVs have been measured. These RVs and the resulting calculations of  $U, V, W$ and $X, Y, Z$, as obtained from D. Rodriguez's kinematics calculator\footnote{http://kinematics.bdnyc.org/}, are listed in Table~\ref{tbl:Summary}. 

\section{Results and Discussion}

\subsection{Colors of GALNYSS stars relative to all {\it Galex}-detected stars in TGAS}

In Fig.~\ref{fig:GvsJmW1} and Fig.~\ref{fig:GvsNUVmG}, we display color-magnitude diagrams (CMDs) for all stars in the TGAS catalog that were detected by {\it Galex}, with a $3''$ match radius used to establish associations between TGAS and {\it Galex} sources. Fig.~\ref{fig:GvsJmW1} is restricted to those {\it Galex}-detected stars in TGAS for which {\it WISE} photometry is also available, where we used a $5''$ match radius to establish associations between TGAS/{\it Galex} and {\it WISE} sources. The stellar main sequence (MS) is evident in both CMDs, as are the horizontal and giant branches (upper panels of Fig.~\ref{fig:GvsJmW1}, \ref{fig:GvsNUVmG}). 

In both CMDs, the Table~\ref{tbl:GaiaData} stars are positioned, for the most part, on the tails of the lower MS distributions, somewhat offset from the expected positions of the latest-type stars (lower panels of Figs.~\ref{fig:GvsJmW1}, \ref{fig:GvsNUVmG}). However, the offsets from the MS lie in different directions in the two CMDs: in the $G$ vs.\ $J-W1$ CMD, the Table~\ref{tbl:GaiaData} stars mostly lie above the lower MS, whereas in the $G$ vs.\ $NUV-G$ CMD (Fig.~\ref{fig:GvsNUVmG}), the Table~\ref{tbl:GaiaData} stars mostly lie to the left of (below) the lower MS. 
This difference is readily understandable in terms of the specific combination of relatively large (pre-main sequence) luminosities and high levels of activity that characterize the stars selected via the criteria for inclusion in GALNYSS \citep{Rodriguez2013}. Specifically, in Fig.~\ref{fig:GvsJmW1}, we see that the absolute $G$ magnitudes of the Table~\ref{tbl:GaiaData} stars are generally smaller than MS stars of similar $J-W1$ color. This offset reflects the fact that absolute $G$ is here serving as a proxy for luminosity ($L_{\rm bol}$), while $J-W1$ provides an effective proxy for photospheric temperature ($T_{\rm eff}$) for K and M stars \citep[e.g.,][]{PecautMamajek2013}. Hence, Fig.~\ref{fig:GvsJmW1} confirms that the GALNYSS subsample in Table~\ref{tbl:GaiaData} is primarily composed of low-mass, pre-MS stars, i.e., stars that lie above MS stars of similarly low $T_{\rm eff}$. Only four Table~\ref{tbl:GaiaData} stars appear to lie on or below the MS locus in Fig.~\ref{fig:GvsJmW1}, and two of these stars, HIP 11152 and TYC 3828-36-1, have previously been identified as young (\S \ref{sec:PrevConsidered}). 

In complementary fashion, Fig.~\ref{fig:GvsNUVmG} demonstrates that the Table~\ref{tbl:GaiaData} stars are chromospherically active. Most of these 19 stars lie to the left of the lower MS in Fig.~\ref{fig:GvsNUVmG} because their levels of chromospheric UV emission are elevated relative to those of late-type MS dwarfs \citep{Rodriguez2013} and, as a consequence, the $NUV-G$ colors of GALNYSS stars are bluer than those of late-type MS stars of similar absolute $G$ mag. Similarly, the $NUV-G$ vs.\ $J-W1$ color-color diagram displayed in Fig.~\ref{fig:NUVmGvsJmW1} shows that the Table~\ref{tbl:GaiaData} stars are bluer---i.e., more chromospherically active---than stars of similar $T_{\rm eff}$ (as reflected in their $J-W1$ colors). This is consistent with the expected active young (pre-MS or ZAMS) star status of the GALNYSS sample as a whole. Figs.~\ref{fig:GvsNUVmG} and \ref{fig:NUVmGvsJmW1} hence show that the {\it Galex} vs.\ {\it Gaia} ($NUV-G$) color, like $NUV-J$ \citep{Rodriguez2013}, serves as  an activity indicator for late-type stars. 

The preceding (color-based) analysis indicates that most, and perhaps all, of the late-type stars in Table~\ref{tbl:GaiaData} are magnetically active, pre-main-sequence (pre-MS) or young MS stars. This conclusion derives additional support from the H$\alpha$ spectral region measurements summarized in Table~\ref{tbl:Obs}: H$\alpha$ appears in emission for most stars, with most stars displaying H$\alpha$ emission EWs in the range $-$0.4 \AA\ to $-$1.4 \AA. This regime of H$\alpha$ emission-line EWs is typical of nearby, young (hence chromospherically active) stars in the mid-K through early-M spectral type range spanned by the objects in Table~\ref{tbl:GaiaData} \citep[see, e.g., Fig.~5 in][]{ZuckermanSong2004}.

We note that none of the Table~\ref{tbl:GaiaData} stars show evidence for the presence of a primordial or debris disk in the form of warm circumstellar dust, in {\it WISE} photometric data. Specifically, late-type stars with dusty disks are expected to have ``red'' $W1-W4$ colors \citep[i.e., $W1-W4 > 1.0$; e.g.,][]{Schneider2012a}, whereas all of the Table~\ref{tbl:GaiaData} stars display $W1-W4<0.4$, with most in the range $0.1 \le W1-W4 \le 0.3$. 

\subsection{Comparison with NYMG members and pre-main sequence isochrones}


In Fig.~\ref{fig:GaiaNYMGs}, we present an absolute $K_S$ vs.\ $G-K_S$ CMD, constructed from {\it Gaia} and {\it 2MASS} data, that serves to compare the optical/near-IR CMD positions of the Table~\ref{tbl:GaiaData} stars with those of members of four particularly well-studied NYMGs: the TW Hya Association (TWA), the $\beta$ Pic Moving Group (Beta Pic), the Tucana-Horologium Association (Tuc Hor) and the AB Dor Assocation (AB Dor). The {\it Gaia} data for these NYMGs were compiled from the photometric ($G$ band) and parallax data available in TGAS for the TWA, Beta Pic, Tuc Hor, and AB Dor member stars listed in \citet[][]{Torres2008}. Specifically, based on the \citet[][]{Torres2008} lists, we find (respectively) 3, 23, 33, and 61 members of the four groups for which TGAS data are available. 

The resulting $K_S$ vs.\ $G-K_S$ CMD is overlaid with pre-MS  isochrones obtained from the models presented in \citet[][top panel of Fig.~\ref{fig:GaiaNYMGs}]{Tognelli2011} and \citet[][bottom panel of Fig.~\ref{fig:GaiaNYMGs}]{Baraffe2015}. The two sets of  model isochrones diverge sharply for values of $G-K_S \ge 2.9$, a color corresponding roughly to spectral type M0 (Figs.~\ref{fig:GaiaNYMGs} and \ref{fig:GaiaIsochrones}). 
It is apparent from Fig.~\ref{fig:GaiaNYMGs} that the TWA, Beta Pic, Tuc Hor, and AB Dor NYMGs present a sequence of generally increasing $K_S$ magnitude (corresponding to decreasing $L_{\rm bol}$) for a given $G-K_S$ color (corresponding to a given $T_{\rm eff}$). This sequence is consistent with the relative ages of these NYMGs, i.e., $\sim$8, $\sim$23, $\sim$45, and $\sim$100--150 Myr, respectively \citep[][and refs.\ therein]{Donaldson2016,Mamajek2016}.  There is considerable scatter, however, some of which is due to the presence of unresolved binaries in the \citet[][]{Torres2008} NYMG membership lists (see \S\ref{sec:Ages}).

It is clear from Fig.~\ref{fig:GaiaNYMGs} that the Table~\ref{tbl:GaiaData} stars for the most part span an age range similar to that of the four NYMGs, i.e., from $\sim$10 Myr (TWA) to $\sim$100 Myr (AB Dor). 
The pre-MS natures of the majority of the Table~\ref{tbl:GaiaData} stars, as well as the approximate age range just noted for these stars, derive further support from both sets of pre-MS isochrones that are overlaid on the data in Fig.~\ref{fig:GaiaNYMGs}.

To further explore this last point we plot, in Fig.~\ref{fig:GaiaIsochrones}, $K_S$ vs.\ $G-K_S$ CMDs in which the \citet[][]{Tognelli2011} and \citet[][]{Baraffe2015} isochrones and the positions of the Table~\ref{tbl:GaiaData} stars are overlaid on the full set of TGAS sources with {\it Galex} and {\it WISE} counterparts (i.e., the same TGAS sample as is illustrated in Figs.~\ref{fig:GvsJmW1}--\ref{fig:NUVmGvsJmW1}). The fact that this MS locus lies along the oldest (80 and 100 Myr) model isochrones (see bottom panel of Fig.~\ref{fig:GaiaIsochrones}) reflects the fact that stars of type K and earlier have nearly settled onto the MS by such ages, but also appears to reinforce the notion that {\it Galex-}selected late-type stars represent a population that is systematically younger than late-type field stars generally \citep{Shkolnik2011}.
Comparing Fig.~\ref{fig:GaiaNYMGs} and the top panel of Fig.~\ref{fig:GaiaIsochrones}, it is readily apparent that the vast majority of the NYMG members and all but four of the Table~\ref{tbl:GaiaData} stars lie above the lower MS locus, as defined by the TGAS sample. 

\subsection{Age estimates and NYMG membership assessments \label{sec:Ages}}

In Fig.~\ref{fig:Good19colorMag} we present a $K_S$ vs.\ $G-K_S$ CMD (as in Fig.~\ref{fig:GaiaNYMGs}) with the axes scaled so as to more easily ascertain the positions of the Table~\ref{tbl:GaiaData} stars relative to the pre-MS isochrones calculated from models in \citet{Tognelli2011} and \citet{Baraffe2015}.
In Table~\ref{tbl:Summary} (column ``Age, CMD''), we list the isochronal ages of individual stars as inferred from these positions in Fig.~\ref{fig:Good19colorMag}. We caution that, for any stars that are unresolved binaries, their absolute magnitudes --- and, therefore, their isochronal (CMD-based) ages --- would be underestimates. In the worst case --- an unresolved binary consisting of components of equal luminosity --- the absolute magnitude could be underestimated by up to 0.75 mag (see below). This would potentially cast doubt on the pre-MS status of some of the Table~\ref{tbl:Summary} stars. Furthermore, the isochronal ages of most Table~\ref{tbl:GaiaData} stars of type M0 and later are uncertain, due to the significant discrepancies between the \citet{Tognelli2011} and \citet{Baraffe2015} model isochrones for values of $G-K_S \ge 2.9$ (Fig.~\ref{fig:Good19colorMag}). For those stars with $G-K_S \ge 2.9$, the \citet[][]{Baraffe2015} isochrones yield younger age estimates than those of \citet[][]{Tognelli2011} for a star of given absolute $K_S$ magnitude and $G-K_S$ color. 

For these reasons, we have also estimated ages via comparison of the available measurements of the EW of the Li {\sc i} $\lambda$6708 absorption line with the empirical trends of Li EW with spectral type illustrated in \citet[][their Fig. 8]{Murphy2015}, for the (15) Table~\ref{tbl:GaiaData} stars for which such Li EW measurements are available. These Li EW-based ages are also listed in Table~\ref{tbl:Summary} (column ``Age, Li''). For the majority of stars, the isochronal (color-magnitude diagram) and Li EW-based ages are in reasonable agreement. In those cases for which the age estimates are discrepant, the  isochronal ages are generally younger than the Li-based ages. Such systematic discrepancies may indicate that the radii of these (pre-MS) stars are larger than predicted by the models; this stellar radial ``inflation'' would be a consequence of the intense magnetic activity characteristic of pre-MS stars \citep[][and references therein]{Jeffries2017,SomersStassun2017}. One must also keep in mind the longstanding problems of an observed spread of Li EWs for stars of given spectral type and age \citep[see, e.g., Fig.~3 in][]{ZuckermanSong2004} and that (presumably coeval) young multiple systems display a spread in Li-based ages \citep{Mamajek2008}. Such dispersion in Li-based age determinations may be due, at least in part, to the apparent dependence of Li depletion on stellar rotational velocity \citep{Messina2016}.  

However, some of the discrepancies  between isochronal and Li-based ages apparent in Table~\ref{tbl:Summary}  may reveal cases in which the presence of a binary companion corrupts the isochronal age estimates.
To illustrate and diagnose the possible effects of binarity on the estimated isochronal ages, we include in Fig.~\ref{fig:Good19colorMag} a (black) curve representing the 80 Myr isochrone from \citet[][]{Tognelli2011} transformed upwards in luminosity by 0.75 mag. This transformed isochrone correponds to the ``worst-case'' scenario of equal-components binary systems whose CMD positions are asymptotically approaching the main sequence. High confidence can be placed in the pre-MS status of the half-dozen Table~\ref{tbl:GaiaData} stars that lie above the black curve in Fig.~\ref{fig:Good19colorMag}, as well as those stars that lie below the black curve for which isochronal ages $<$100 Myr are supported by Li absorption line EW measurements. Cases of suspected or confirmed binaries --- some of which are indicated by the aforementioned discrepancies between isochronal and Li-based ages --- are noted for individual systems in \S\ \ref{sec:Individuals}. Cases in which a system's possible binarity would likely rule out young star status (e.g., HIP 59748) are noted in \S\ \ref{sec:Imposters}.

We have used the isochronal and Li-based age estimates in combination with comparisons of the $U, V, W$ and $X, Y, Z$ values for individual stars (\S\ 2.4; Table~\ref{tbl:Summary}) with the loci of space velocities and positions of known NYMGs \citep[e.g.,][and references therein]{Torres2008,MaloPlus2014,Mamajek2016} to evaluate the potential NYMG membership of individual Table~\ref{tbl:GaiaData} stars. We find that most Table~\ref{tbl:GaiaData} stars whose ages are commensurate with membership in a given NYMG(s) are found far from those NYMGs in $UVW$ and/or $XYZ$ space. The few cases of stars that have potential NYMG associations are noted in the last column of Table~\ref{tbl:Summary} and are discussed in \S\  \ref{sec:Individuals}. We note that the BANYAN II on-line NYMG membership probability calculator\footnote{http://www.astro.umontreal.ca/~gagne/banyanII.php} \citep{Gagne2014} appears to support these potential NYMG associations in only one case (i.e., HIP 3556; see \S\ \ref{sec:PrevConsidered}).

\section{Individual Objects \label{sec:Individuals}}

\subsection{Stars not previously considered as young}

To the best of our knowledge, nine Table~\ref{tbl:GaiaData} stars have not previously been mentioned, listed, or otherwise included in published investigations concerned with NYMGs or individual nearby, young stars. The majority of these are found in the northern sky (\S \ref{sec:NorthernStars}). Two of the nine Table~\ref{tbl:GaiaData} objects not previously considered as young lie a mere $\sim$20 pc from Earth, but both may be unresolved binaries whose components are in fact young MS (as opposed to pre-MS) stars (\S \ref{sec:Imposters}).

\subsubsection{Newly identified nearby, young stars in the northern sky \label{sec:NorthernStars}} 

The isochronal ages of $<$40 Myr we infer for five Table~\ref{tbl:GaiaData} objects that are considered here for the first time as young --- TYC 3824-657-1, TYC 4632-1171-1, TYC 3462-1056-1, TYC 2627-594-1, and TYC 4448-2206-1  --- are surprising given their positions in the northern sky, far from the nuclei of known NYMGs. Two of these stars, TYC 4632-1171-1 and TYC 4448-2206-1, lie north of +70$^\circ$.  
The youth of these stars is reflected in their large values of $\log{(L_X/L_{\rm bol})}$ (Table~\ref{tbl:DerivedData}) and, in four of five cases (all but TYC 4448-2206-1), the detection of Balmer line emission \citep[Table~\ref{tbl:Obs};][]{Dragomir2007}.  

Among these stars, only
TYC 4632-1171-1 displays Li absorption (Fig.~\ref{fig:sampleSpectra}). Based on its Li absorption line depth, this star appears to be significantly older than inferred from comparison with isochrones (Table~\ref{tbl:Summary}). Its space velocity is not far from that of the AB Dor $UVW$ locus \citep{Torres2008}, and the upper end of the age range inferred from its Li EW would be consistent with AB Dor membership. However, it lies $\sim$40 pc from the nucleus of that group. 

The ages estimated from nondetections of Li absorption in the other four stars are also discrepant with (older than) their isochronal ages. This suggests all five stars are binary systems that are unresolved by {\it Gaia}. Of the four, only TYC 3462-1056-1 is a known binary, although its M2 companion, at a separation of 1.8$''$ (170 AU), does not affect the inferred isochronal age of the primary \citep{Janson2012}. 
Even if TYC 3824-657-1, TYC 4448-2206-1, TYC 2627-594-1 and the primary of TYC 3462-1056-1 were equal-components binaries, however, their isochronal ages would still be $\stackrel{<}{\sim}$80 Myr (Fig.~\ref{fig:Good19colorMag}). 

On the other hand, the space velocities of TYC 2627-594-1 and TYC 4448-2206-1 are highly anomalous among known nearby young stars \citep[][]{Torres2008,Mamajek2016}. It is possible these stars are runaways from southern NYMGs. If not, there is reason to doubt their pre-MS status, despite the young ages inferred from comparison with isochrones.

\subsubsection{TYC 5708-357-1: young star without a home?}

The star TYC 5708-357-1 displays moderate H$\alpha$ emission and Li absorption lines in our Lick/Hamilton spectroscopy. The isochronal age of TYC 5708-357-1, $\sim$40 Myr (Fig.~\ref{fig:Good19colorMag}), is consistent with its Li {\sc i} $\lambda$6708 line EW ($\sim$120 m\AA; Table~\ref{tbl:Summary}) assuming our ($G-K_S$ color-based) classification of mid-K is accurate. However, while it is not found as far from known NYMGs as the northern-sky stars just discussed, the $UVW$ of TYC 5708-357-1 are inconsistent with membership in known NYMGs with ages $\sim$40 Myr \citep[][]{Mamajek2016}. 

\subsubsection{Stars with isochronal ages $\stackrel{<}{\sim}$80 Myr that may be unresolved, main-sequence binaries \label{sec:Imposters}}

{\it HIP 59748 (= BD +49 2126) and TYC 3529-1437-1.} Due to their proximity ($D \approx 20$ pc), these two stars have been included in numerous catalogs and papers devoted to nearby, bright, late-type dwarfs \citep[e.g.,][]{Frith2013}. However, neither star had previously been flagged as particularly young. Although both stars have isochronal ages of $\le$100 Myr (Fig.~\ref{fig:Good19colorMag}; Table~\ref{tbl:Summary}), it appears likely that these isochronal ages are underestimates, due to binarity. 

The isochronal age of TYC 3529-1437-1 is uncertain even if the star is single, mainly due to the discrepancy between model isochrones for its regime of $G-K_s$. Our nondetection of Li absorption would not be inconsistent with the younger end of the isochronal age range we infer from Fig.~\ref{fig:Good19colorMag} ($\sim$40--100 Myr). The star's space velocity is similar to the mean $UVW$ of the $\sim$45 Myr-old Carina group, and not far from the $UVW$ of $\sim$100 Myr-old AB Dor \citep{Mamajek2016}. Its distance and Galactic ($XYZ$) position is a better match to the latter NYMG and, hence, TYC 3529-1437-1 would appear to be more viable as a AB Dor candidate. However, while BANYAN II yields a 64\% probability that TYC 3529-1437-1 is a ``young'' (age $<$1 Gyr) field star given the star's {\it Gaia} data (Table~\ref{tbl:GaiaData}) and RV (Table~\ref{tbl:Summary}) as input, the BANYAN II membership probability for AB Dor is only 3\%.

The young-star status of the early-M star HIP 59748 is also uncertain. While its isochronal age is only $\sim$80 Myr (Fig.~\ref{fig:Good19colorMag}), comparison with models of late-type stars yields an age of a few Gyr \citep[][]{Mann2015}. As is the case for the early-M TYC 3529-1437-1, our Li absorption line upper limit is not particularly constraining. The $UVW$ velocities of HIP 59748 do not match those of any known NYMG and, while the isochronal age we determine is compatible with membership in AB Dor, it lies 20 pc from this NYMG in $Z$. While HIP 59748 was not found to be a binary in the high-resolution imaging survey of \citet[][]{Jodar2013}, the star was flagged as a potential proper motion binary by \citet{Frankowski2007}, and we find a significant discrepancy between its catalogued RV ($-$13.24$\pm$0.24 km s$^{-1}$) and the RV we measure from our Keck/ESI spectrum ($-$19.4$\pm$1.5 km s$^{-1}$). We further note that an age as young as $\sim$80 Myr is difficult to reconcile with the star's nondetection in the RASS (Table~\ref{tbl:DerivedData}). Based on the resulting inferred upper limit of $\log{L_X/L_{\rm bol}} < -4.4$, this star is far weaker in X-rays than typical of early-M stars in young clusters \citep[e.g., ][]{Kastner1997,Kastner2016} --- an observation consistent with the fact that HIP 59748 is among the weaker UV excess sources among the Table~\ref{tbl:GaiaData} stars and displays relatively weak, variable H$\alpha$ emission \citep[literature values of H$\alpha$ EW range from $-$0.36 to +0.39 \AA;][]{Lepine2013, Gaidos2014}. 

{\it TYC 3475-768-1.}
The strength of TiO bands in our Lick/Kast spectrum and its $G-K_S$ color are consistent with the K7 spectral type originally determined for TYC 3475-768-1 by \citet{Stephenson1986a}. The Li absorption line upper limit we infer from our Lick/Kast spectroscopy (EW $<$90 m\AA; Table~\ref{tbl:Obs}) implies this star is older than its isochronal age of $\sim$35 Myr. As in the cases of HIP 59748 and TYC 3529-1437-1, if TYC 3475-768-1 is an equal-components binary, its isochronal age would be $>$80 Myr (Fig.~\ref{fig:Good19colorMag}). Its $UVW$ space velocity is inconsistent with known NYMGs and nearby young stars generally.

\subsection{Stars Previously Considered as Young \label{sec:PrevConsidered}}

Ten Table~\ref{tbl:GaiaData} stars are either confirmed or candidate members of NYMGs, or have been considered and rejected as members of specific NYMGs. Seven of these (i.e., all except HIP 3556, HIP 11152, and TYC 3828-36-1) were included in the catalog of candidate nearby young stars compiled by \citet{Torres2006}, but were not among the candidate $\beta$PMG members identified as part of that study.  Three of these ten stars are components or potential components of wide binaries (\S \ref {sec:WideBinaries}). 

\subsubsection{Components (and potential components) of wide binaries \label{sec:WideBinaries}} 

{\it TYC 8246-2900-1.} 
The combination of $UVW$ velocities and $\sim$20 Myr isochronal age we infer for the early-K star TYC 8246-2900-1 suggests that this star could be a member of the $\beta$PMG. However, the weak Li absorption measured for TYC 8246-2900-1, which is indicative of an age $>40$ Myr, argues against this inference. This discrepancy between Li-based and isochronal ages is indicative of a binary nature for TYC 8246-2900-1. It is furthermore intriguing that TYC 8246-2900-1 is found $\sim$8.3$'$ from the well-studied, dusty, young F-type binary system  HD 113766, \citep[][and refs.\ therein]{Lisse2017}, given that the proper motions of the two are consistent within the errors \citep[TYC 8246-2900-1: $\mu_\alpha = -$33.2$\pm$1.5 mas yr$^{-1}$,  $\mu_\delta= -$19.0$\pm$0.4 mas yr$^{-1}$, Table~\ref{tbl:GaiaData}; HD 113766: $\mu_\alpha = -$34.09$\pm$0.82 mas yr$^{-1}$, $\mu_\delta = -$17.90$\pm$0.59 mas yr$^{-1}$,][]{vanLeeuwen2007}. This raises the possibility that the two stars constitute a comoving pair with projected separation $\sim$47 kAU. If so, the HD 113766 plus TYC 8246-2900-1 system would join the ranks of hierachical binaries consisting of dusty, intermediate-mass primaries and wide-separation, late-type secondaries \citep[][and refs.\ therein]{Kastner2012,Zuckerman2015}. 

However, the RV of TYC 8246-2900-1 \citep[+17.3 km s$^{-1}$][]{Torres2006} differs significantly from that of HD 113766 (listed in Simbad as $-$1.0$\pm$3.0), and the Hipparcos parallax of the latter \citep[$\pi = $8.16$\pm$1.01 mas;][]{vanLeeuwen2007} differs by 2.5$\sigma$ from the {\it Gaia} parallax of the former ($\pi = $10.56$\pm$0.32 mas; Table~\ref{tbl:GaiaData}). The potential close binary nature of TYC 8246-2900-1 might explain the discrepancy in RVs. Radial velocity monitoring of TYC 8246-2900-1 as well as the forthcoming {\it Gaia} parallax distance to HD 113766 should provide tests of the hypothesis that HD 113766 and TYC 8246-2900-1 represent a wide binary. If TYC 8246-2900-1 and HD 113766 are indeed physically associated, this would imply that the age of the latter is also at least 20 Myr and likely $>$40 Myr, which is considerably older than previously determined \citep[][and refs.\ therein]{Lisse2017}.

{\it TYC 9114-1267-1.} This star is the secondary component of the 26$''$ (834 AU) projected separation binary WDS 21214$–$6655, whose primary, V390 Pav  (= HD 202746),  is a candidate Tuc-Hor member \citep{Zuckerman2001} of spectral type G9 \citep{Elliott2016}. Our DuPont/BC832 spectroscopy supports the classification of TYC 9114-1267-1 as K7 \citep{Torres2006}. 
The ($\sim$40 Myr) age we estimate for TYC 9114-1267-1 and our Li absorption line upper limit ($<$60 m\AA) are consistent with Tuc-Hor membership for the V390 Pav plus TYC 9114-1267-1 system (WDS 21214$–$6655) and appears to rule out $\beta$PMG membership for this system \citep[see also][]{Schlieder2010,Alonso-Floriano2015}. On the other hand, while the TGAS parallaxes of V390 Pav ($\pi = $31.24$\pm$0.26 mas) and TYC 9114-1267-1 ($\pi = $31.1$\pm$0.8 mas) agree to within the errors, their TGAS proper motions differ significantly (V390 Pav: $\mu_\alpha = $95.66$\pm$0.06, $\mu_\delta= -$101.20$\pm$0.09 mas yr$^{-1}$; TYC 9114-1267-1: $\mu_\alpha = $116.3$\pm$1.2, $\mu_\delta= -$90.9$\pm$1.6 mas yr$^{-1}$), as do their RVs \citep[V390 Pav: $-$24.1 km s$^{-1}$; TYC 9114-1267-1: +3.3 km s$^{-1}$;]{Torres2006}. This discrepancy in space motions is especially surprising given the longstanding classification of V390 Pav and TYC 9114-1267-1 as the comoving binary system WDS 21214$–$6655.\footnote{ \citet{Skiff2014} notes this system as a ``common proper motion pair'' based on analysis of images dating back to the plates of \citet{CannonPickering1924}.}

{\it TYC 9129-1092-1.} This late-K star has a fainter (early-M) companion, \\ 2MASS~J23351994$-$6433223, at a projected separation of $\sim$22.36$''$ (1214 AU) and position angle $\sim$340$^\circ$. The two evidently constitute a common (high) proper motion pair \citep{Finch2007}. Although TYC 9129-1092-1 presents a reasonable $XYZ$ and age match to the $\beta$PMG, the star's $UVW$ velocities appear anomalous for nearby, young stars \citep[][]{Mamajek2016}. 

\subsubsection{Other Stars}

{\it HIP 3556.}
This M-type star has been proposed as a Tuc-Hor Association member \citep{Zuckerman2001,PecautMamajek2013}, and BANYAN II yields a 99.5\% probability that HIP 3556 is a member of Tuc-Hor given the input position, parallax, proper motion, and RV data from Tables~\ref{tbl:GaiaData} and \ref{tbl:Summary}.  In our MPG/FEROS spectroscopy, HIP 3556 appears as a double-lined spectroscopic binary (Table~\ref{tbl:Obs}) so, in calculating its $UVW$ space velocity, we have adopted the RV measured by \citet{Torres2006}.  The isochronal age range listed in Table~\ref{tbl:Summary}, 40--80 Myr, is an estimate assuming the components of HIP 3556 contribute equally to its absolute $K_S$ mag. The younger end of this range, and the star's weak H$\alpha$ emission and lack of detectable Li absorption, is consistent with membership in Tuc-Hor \citep[age $\sim$45 Myr; see][and refs.\ therein, and Fig~\ref{fig:GaiaNYMGs}]{Murphy2015,Mamajek2016}. 

{\it HIP 11152.}
This star is a candidate member of the $\sim$20 Myr-old $\beta$PMG \citep{Schlieder2012}, and BANYAN II confirms that its space velocity and Galactic position are compatible with $\beta$PMG membership at the $\sim$90\% confidence level. However, its $\beta$PMG membership status was questioned on the basis of its combination of early-M spectral type and lack of Li absorption \citep{MaloPlus2014}. While our IRTF/SpeX spectrum 
confirms the M1 spectral type determined for HIP 11152 by \citet{Schlieder2012} and \citet{PecautMamajek2013}, the isochronal age we infer, $\sim$80--100 Myr (Fig.~\ref{fig:Good19colorMag}), and the star's somewhat discrepant $UVW$, appears to cast further doubt on HIP 11152's membership in the $\beta$PMG. Indeed, this star appears to be a good $XYZ$ and age match to AB Dor, although its space velocity is far from the AB~Dor $UVW$ locus \citep{Torres2008}.


{\it TYC 6022-1079-1 and TYC 3828-36-1 (= GJ 3653).} The $UVW$ space velocities and isochronal ages inferred here for these two stars  (Table~\ref{tbl:Summary}) are suggestive of membership in the Columba Association \citep[see][and refs.\ therein]{Mamajek2016}. The Li {\sc i} $\lambda$6708 EW of TYC 6022-1079-1 \citep[120 m\AA;][]{Torres2006} is also consistent with Columba membership. The star TYC 3828-36-1 was previously identified as a candidate nearby, young (age $<$150 Myr) early-M star on the basis of excess UV emission by \citet{Shkolnik2009}, and was confirmed as a {\it bona-fide} UV-excess young star candidate (as opposed to, e.g., a tidally interacting binary system) by \citet{Ansdell2015}. Here, we refine the age estimate for TYC 3828-36-1 to the range 40--60 Myr (Fig~\ref{fig:Good19colorMag}); the younger end of this range would be consistent with membership in Columba. However, the Galactic positions of TYC 6022-1079-1 and TYC 3828-36-1 place both stars far from the Columba $XYZ$ locus. TYC 6022-1079-1 appears to sit above rather than below the plane  (i.e., it lies at positive rather than negative $Z$), unlike the majority of Columba members, while TYC 3828-36-1 --- like the stars discussed in \S \ref{sec:NorthernStars} --- lies high in the northern sky, far from known NYMGs. BANYAN II yields no probability of Columba membership for either star.

{\it TYC 8246-1527-1.} Although this star lies near the aforementioned TYC 8246-2900-1 in the sky and is essentially identical in $\mu_\delta$, it differs significantly in distance and $\mu_\alpha$, which indicates that the two systems are not bound. Furthermore, they have contrasting isochronal ages ($\sim$20 Myr for TYC 8246-2900-1 vs.\  $\sim$10 Myr for TYC 8246-1527-1; Fig.~\ref{fig:Good19colorMag}), consistent with their contrasting Li {\sc i} $\lambda$6708 EWs (80 m\AA\ and 470 m\AA, respectively; Table~\ref{tbl:Summary}). Given its position in the sky and our estimated age, TYC 8246-1527-1 may be a previously unidentified member of the $\sim$8 Myr-old TW Hya Association (TWA). Its $UVW$ velocities (Table~\ref{tbl:Summary}) are also marginally consistent with TWA membership \citep[e.g.,][]{Mamajek2016}.
However, its distance of $\sim$$110$ pc is approximately twice the median $D$ of known TWA members, and BANYAN II yields 0\% probability of TWA membership. Hence, it is more likely that TYC 8246-1527-1  is a member of the vast Sco-Cen young star complex \citep[e.g.,][and refs.\ therein]{PecautMamajek2016}. 


{\it TYC 7313-1015-1 and HIP 114252 (= HK Aqr).} Both of these stars' space velocities are atypical for nearby, young stars \citep[][]{Mamajek2016}, although the $XYZ$ and age of HIP 114252 appear consistent with membership in AB Dor. TYC 7313-1015-1 is the only star in Table~\ref{tbl:GaiaData} whose Li-based age (30--50 Myr) would make it younger than its isochronal age (80 Myr). Additional spectroscopic observations of these stars is warranted, to ascertain whether they are spectroscopic binaries and to investigate spectral diagnostics of youth. 

\section{Conclusions}

\begin{figure*}[!h]
\centering
\includegraphics[height=3.5in,angle=0]{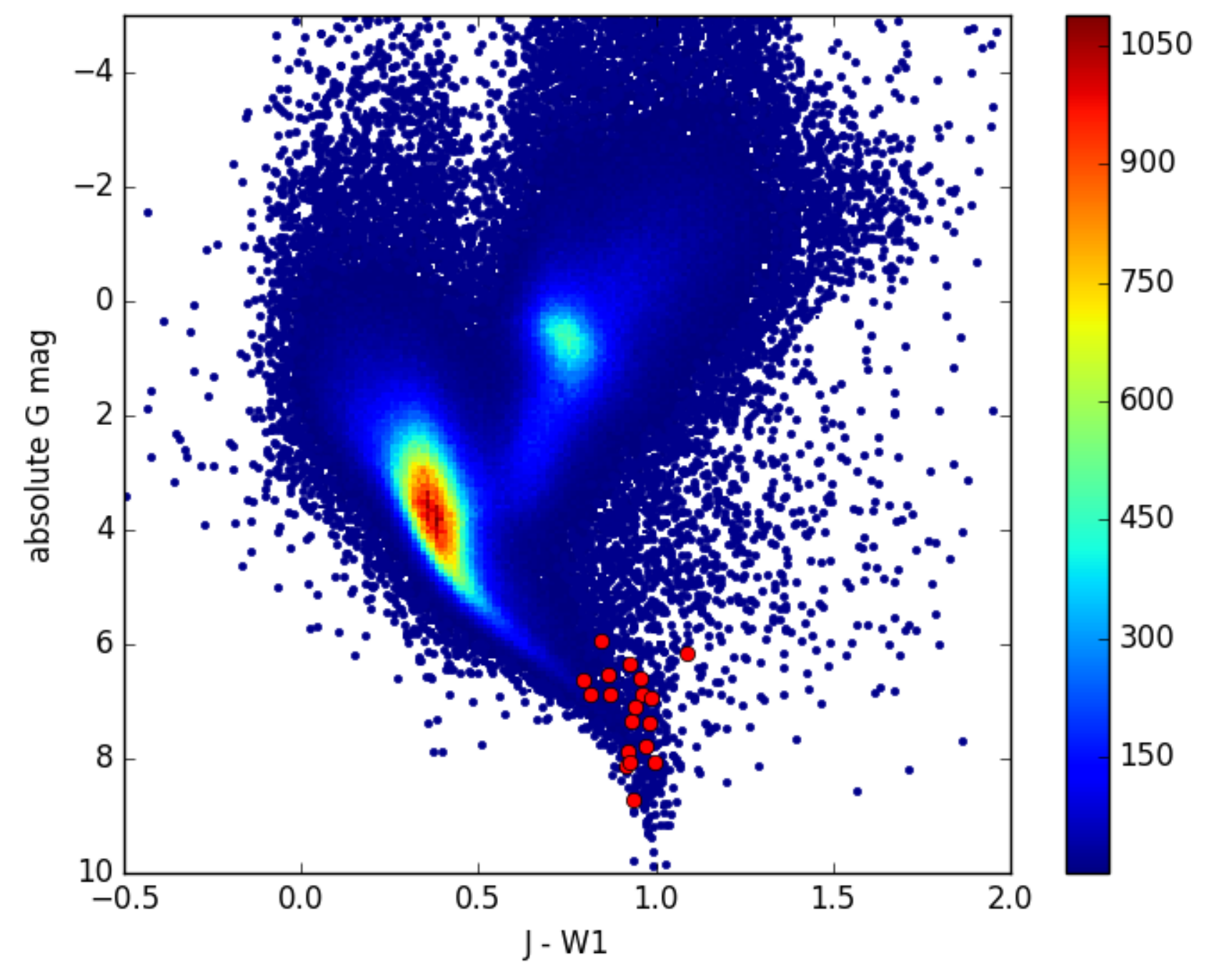}
\includegraphics[height=3.5in,angle=0]{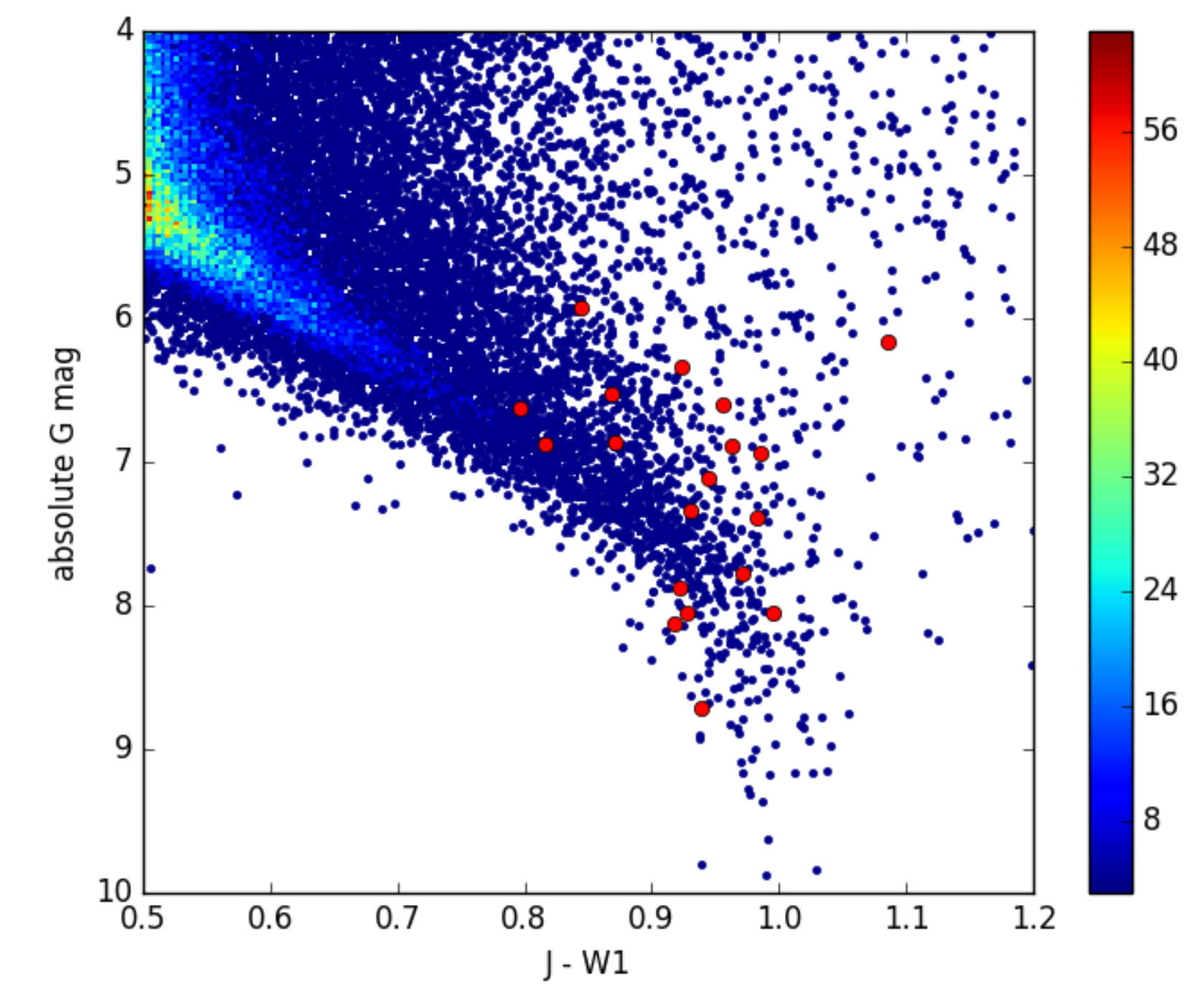}
\caption{\footnotesize 
Color-magnitude diagram (CMD) of  absolute $G$ vs.\ $J-W1$ for {\it WISE}- and {\it Galex}-detected objects in the {\it Gaia} TGAS catalog, plotted in the form of a density image (see color bar for density to color mapping), with the positions of the 19 Table 1 stars indicated as filled red circles. In the top panel, the stellar main sequence (MS) appears as a well-defined locus of high density extending diagonally down and to the right from $G \sim 0$ to $G \sim 9$; the horizontal and giant branches are also apparent  above the lower MS, extending diagonally upwards from $J-W1 \sim 0.5$ to $J-W1 \sim 1.5$ at absolute $G$ magnitudes $< 4$. The CMD in the bottom panel displays a zoomed-in version of the CMD in the top panel that is centered on the lower MS.}
\label{fig:GvsJmW1}
\end{figure*}

\begin{figure*}[!h]
\centering
\includegraphics[height=3.5in,angle=0]{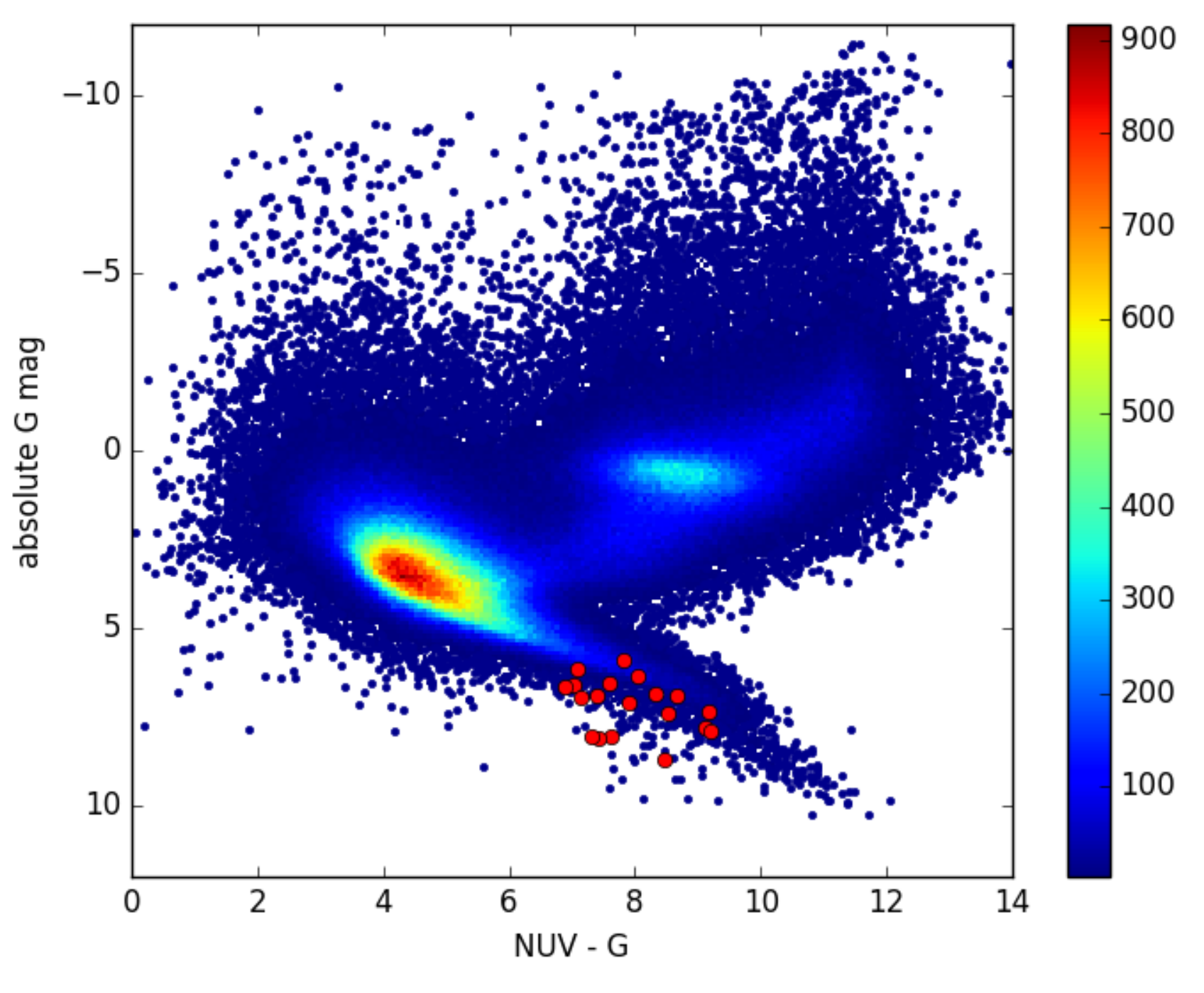}
\includegraphics[height=3.5in,angle=0]{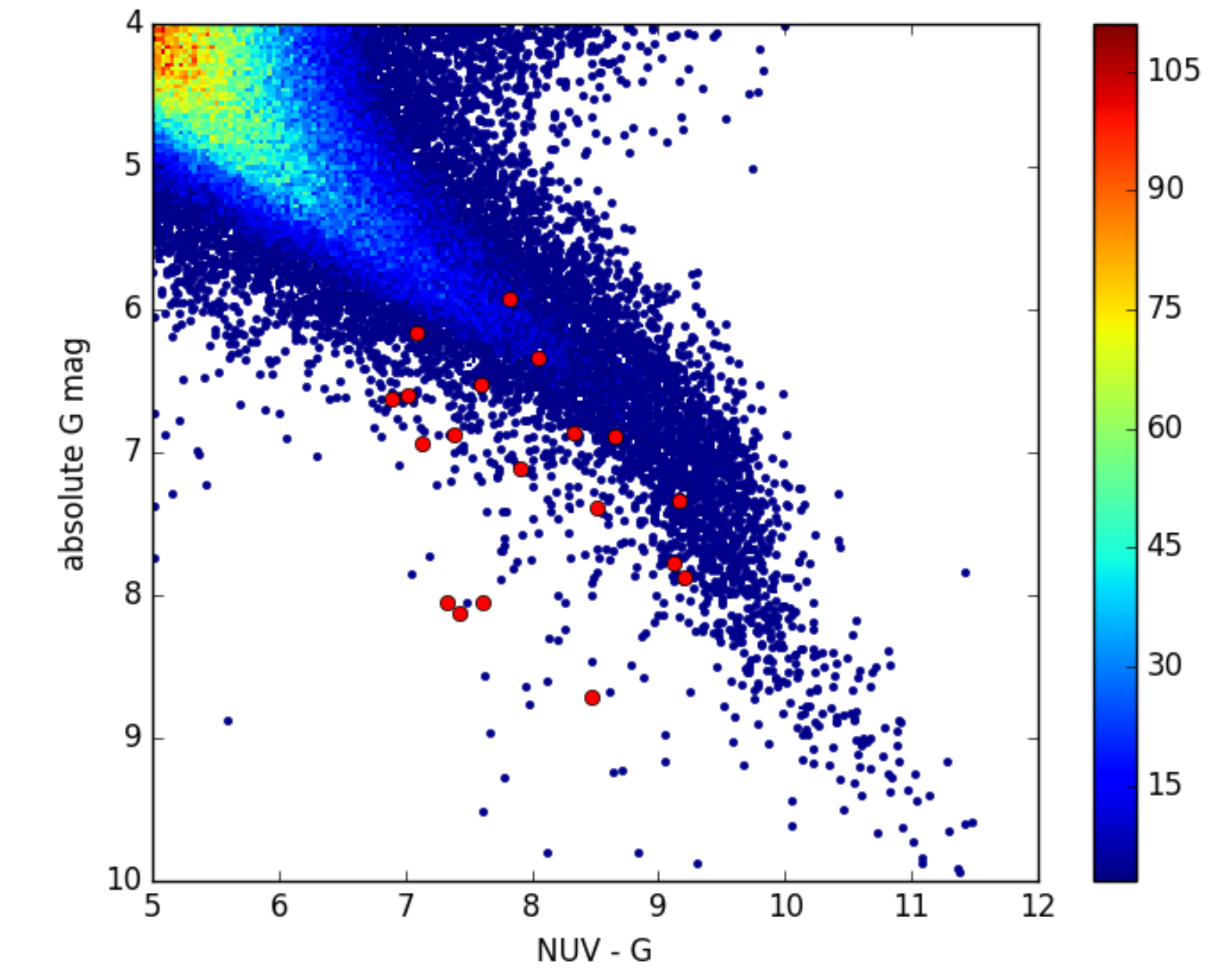}
\vspace{-.18in}
\caption{\footnotesize 
CMD of absolute $G$ vs.\ $NUV-G$ for all {\it Galex}-detected objects in the {\it Gaia} TGAS catalog, rendered as in Fig.~\ref{fig:GvsJmW1},  with the positions of the 19 Table 1 stars indicated as filled red circles. The loci of MS stars and giants are again well-defined. The CMD in the bottom panel displays a zoomed-in version of the CMD in the top panel that is centered on the lower MS.}
\label{fig:GvsNUVmG}
\end{figure*}

\begin{figure*}[!h]
\centering
\includegraphics[height=3.5in,angle=0]{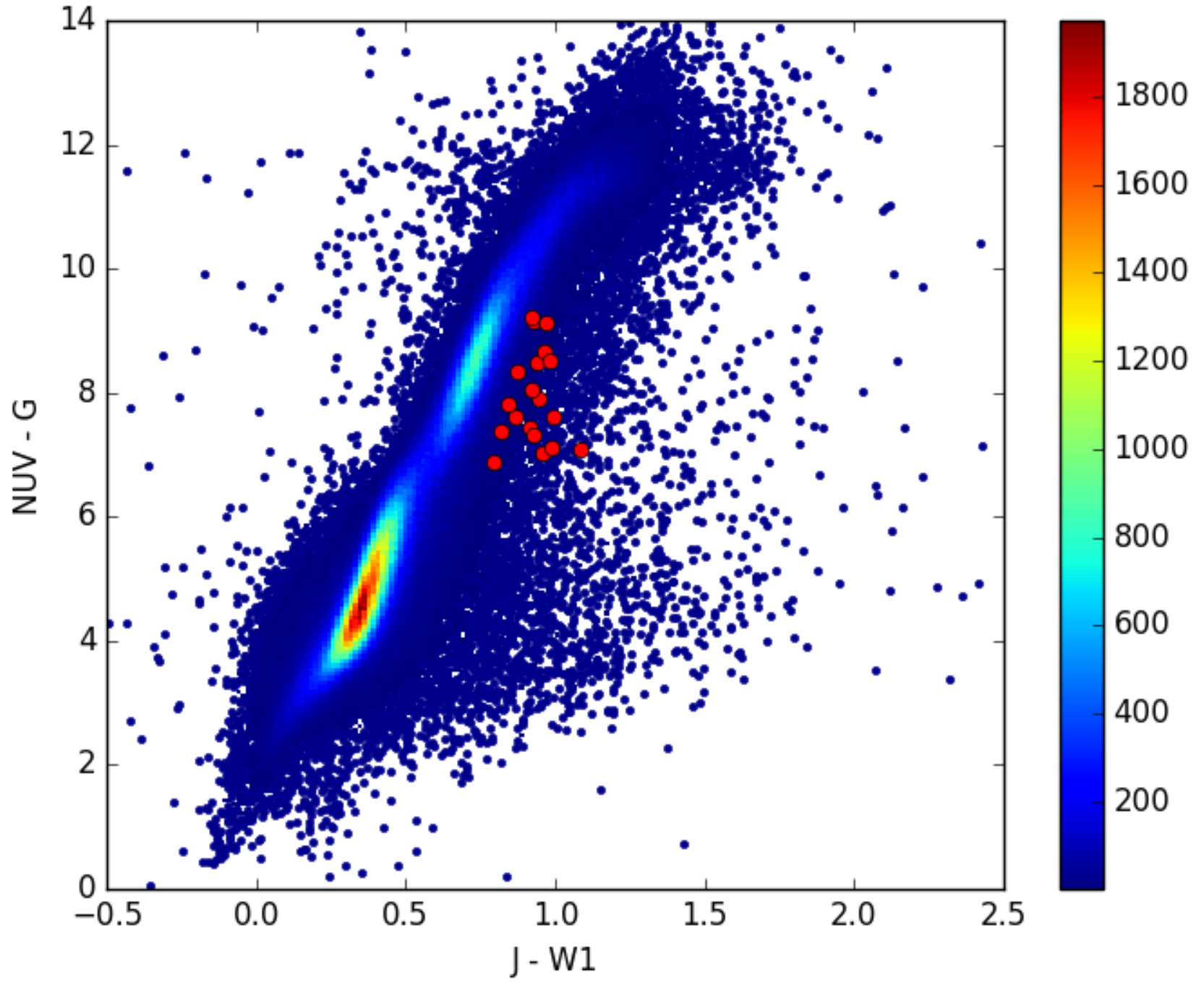}
\includegraphics[height=3.5in,angle=0]{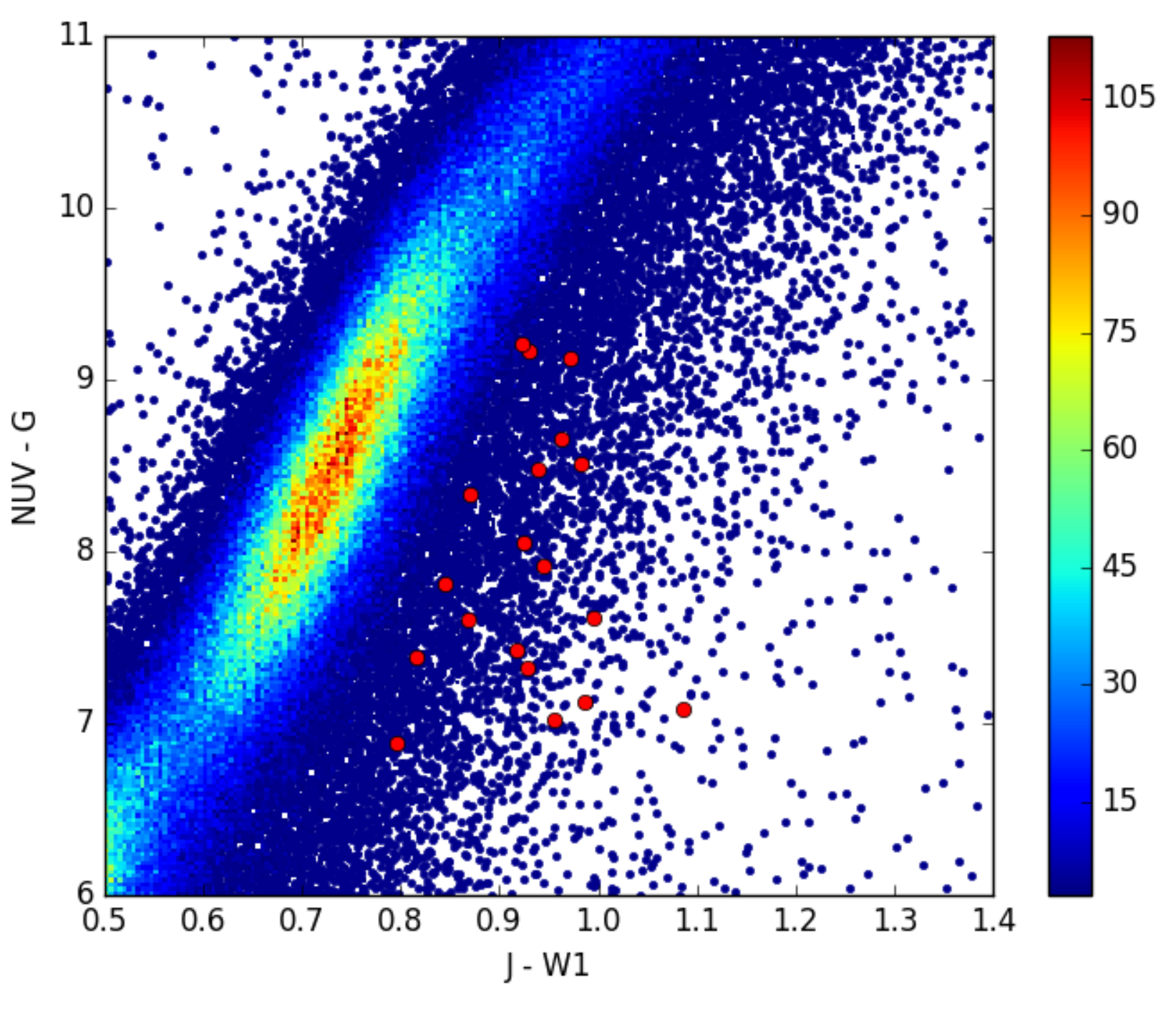}
\caption{\footnotesize 
Color-color diagram of  $NUV-G$ vs.\ $J-W1$ for {\it WISE}- and {\it Galex}-detected objects in the {\it Gaia} TGAS catalog, rendered as in Figs.~\ref{fig:GvsJmW1}, ~\ref{fig:GvsNUVmG}. The positions of the 19 Table 1 stars are indicated as filled red circles.}
\label{fig:NUVmGvsJmW1}
\end{figure*}

We have analyzed {\it Tycho} {\it Gaia} Astrometric Solution (TGAS) catalog parallax and proper motion data newly available in {\it Gaia} Data Release 1 for a subsample of candidate nearby, young stars identified as part of the {\it Galex} Nearby Young Star Survey \citep[GALNYSS;][]{Rodriguez2013}, with the aim of confirming the young-star status of these stars, ascertaining their approximate ages, and assessing potential membership in nearby young moving groups (NYMGs). Of the 50 GALNYSS stars for which parallax and proper motion data are available in TGAS, 19 have parallax distances $\le$120 pc (Table~\ref{tbl:GaiaData}), and our analysis is restricted to these 19 stars (Tables~\ref{tbl:DerivedData}--\ref{tbl:Summary}). The youth of most  of these 19 stars is readily evident from their positions, relative to the loci of main-sequence stars and giants, in color-magnitude and color-color diagrams constructed for all {\it Galex}- and {\it WISE}-detected stars in TGAS (\S 3.1; Figs.~\ref{fig:GvsJmW1}--\ref{fig:NUVmGvsJmW1}). 

\begin{figure*}[!h]
\centering
\includegraphics[height=3.5in,angle=0]{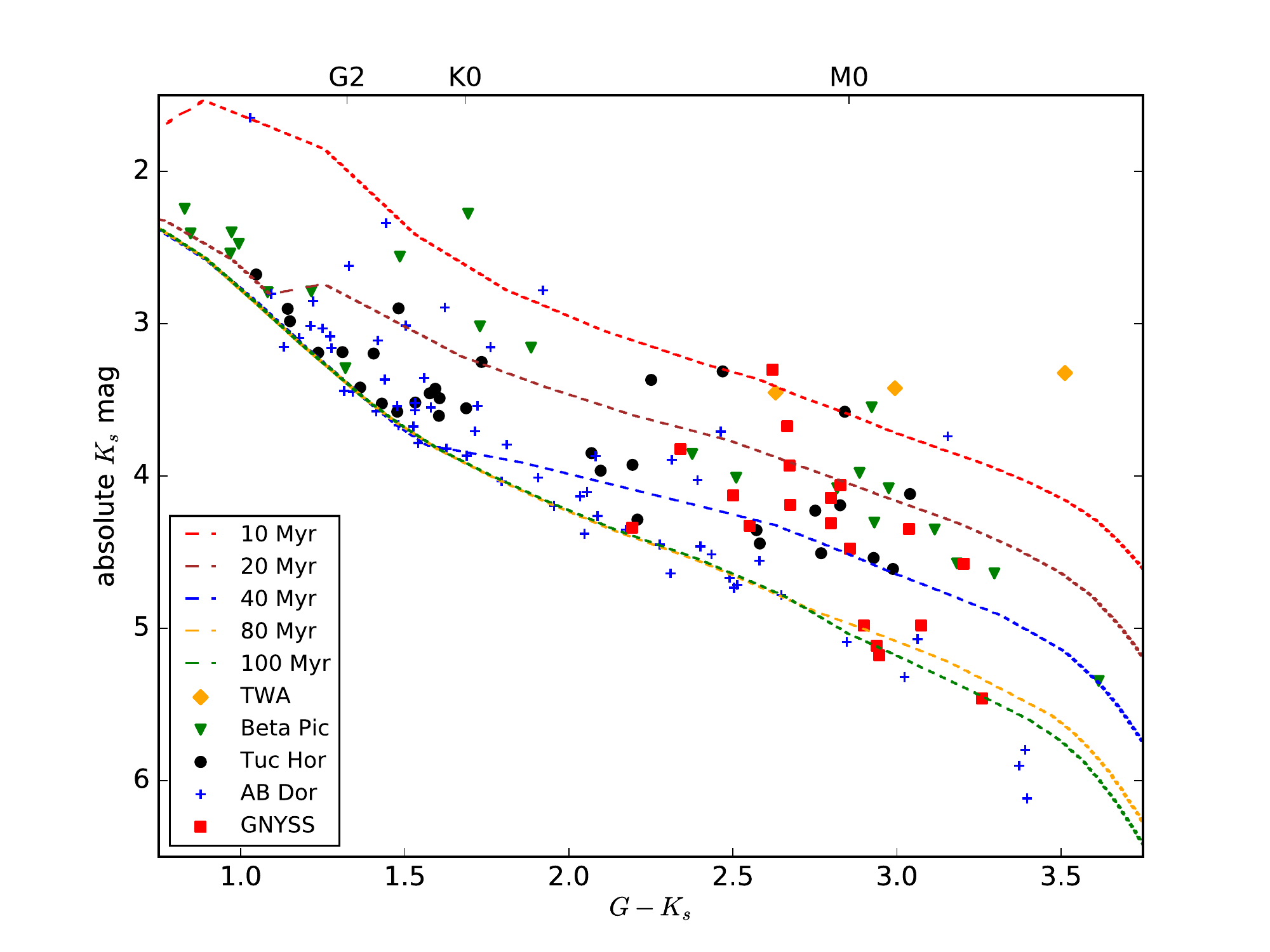}
\includegraphics[height=3.5in,angle=0]{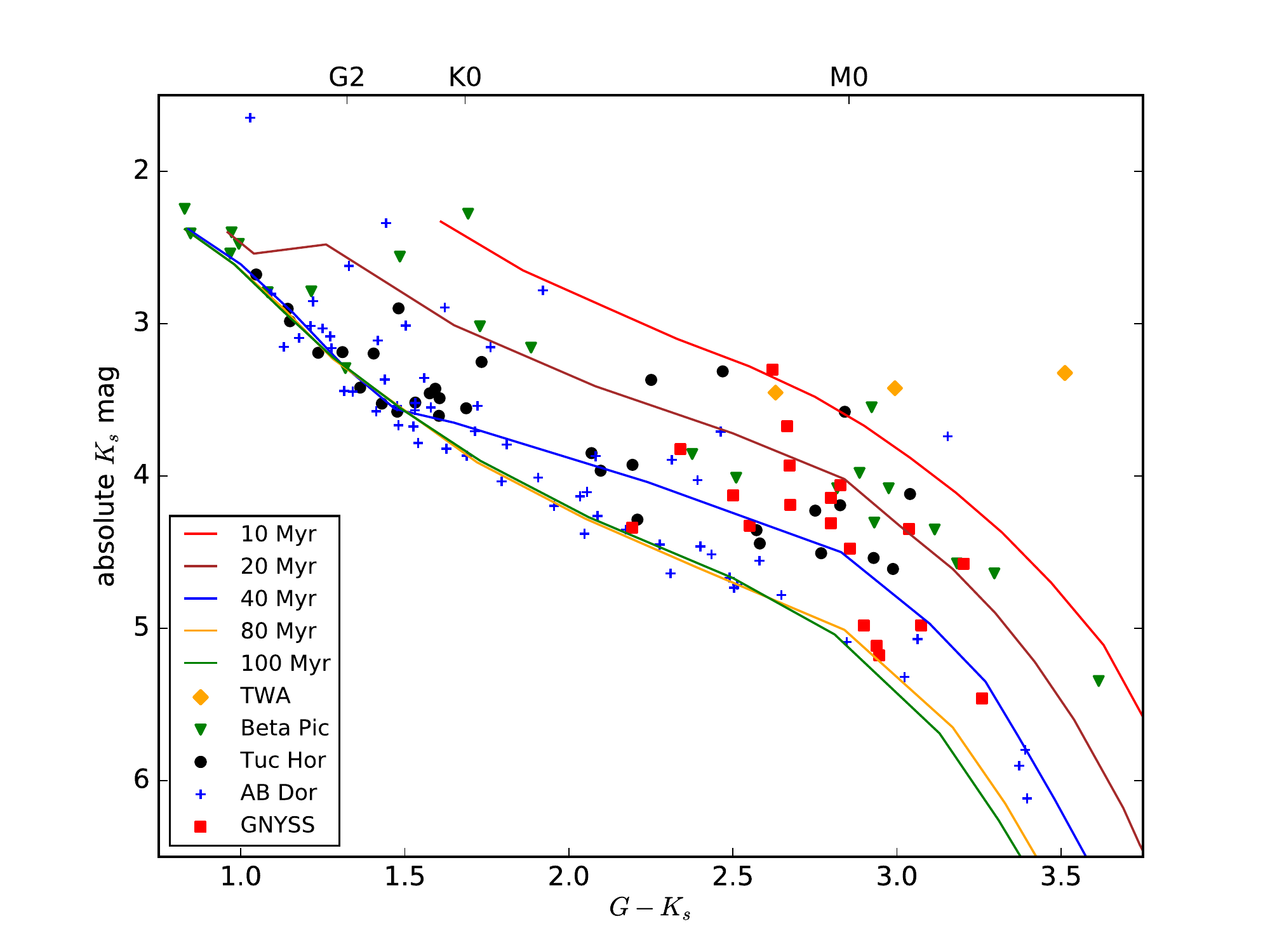}
\vspace{-.18in}
\caption{\footnotesize 
Absolute $K_S$ vs.\ $G-K_S$ CMD for members of four well-established NYMGs \citep[based on membership lists in][where symbols for specific NYMG members are as indicated in legend]{Torres2008} and the Table~\ref{tbl:GaiaData} stars (red squares). The dashed and solid curves in the top and bottom panels, respectively, are pre-main sequence isochrones obtained from models presented in \citet{Tognelli2011} and \citet{Baraffe2015} for ages of 10 Myr  (red; top isochrones), 20 Myr   (brown), 40 Myr   (blue), 80 Myr   (orange), and 100 Myr   (green; bottom isochrones). The synthetic $G$ magnitudes used in these isochrones were computed from the model atmosphere $V$ and $I$ magnitudes using the {\it Gaia}  pre-launch relations presented in \citet{Jordi2010}. The spectral type sequence indicated along the top of each frame is derived from the model atmosphere $T_{\rm eff}$ values.}
\label{fig:GaiaNYMGs}
\end{figure*}

\begin{figure*}[!h]
\centering
\includegraphics[height=3.5in,angle=0]{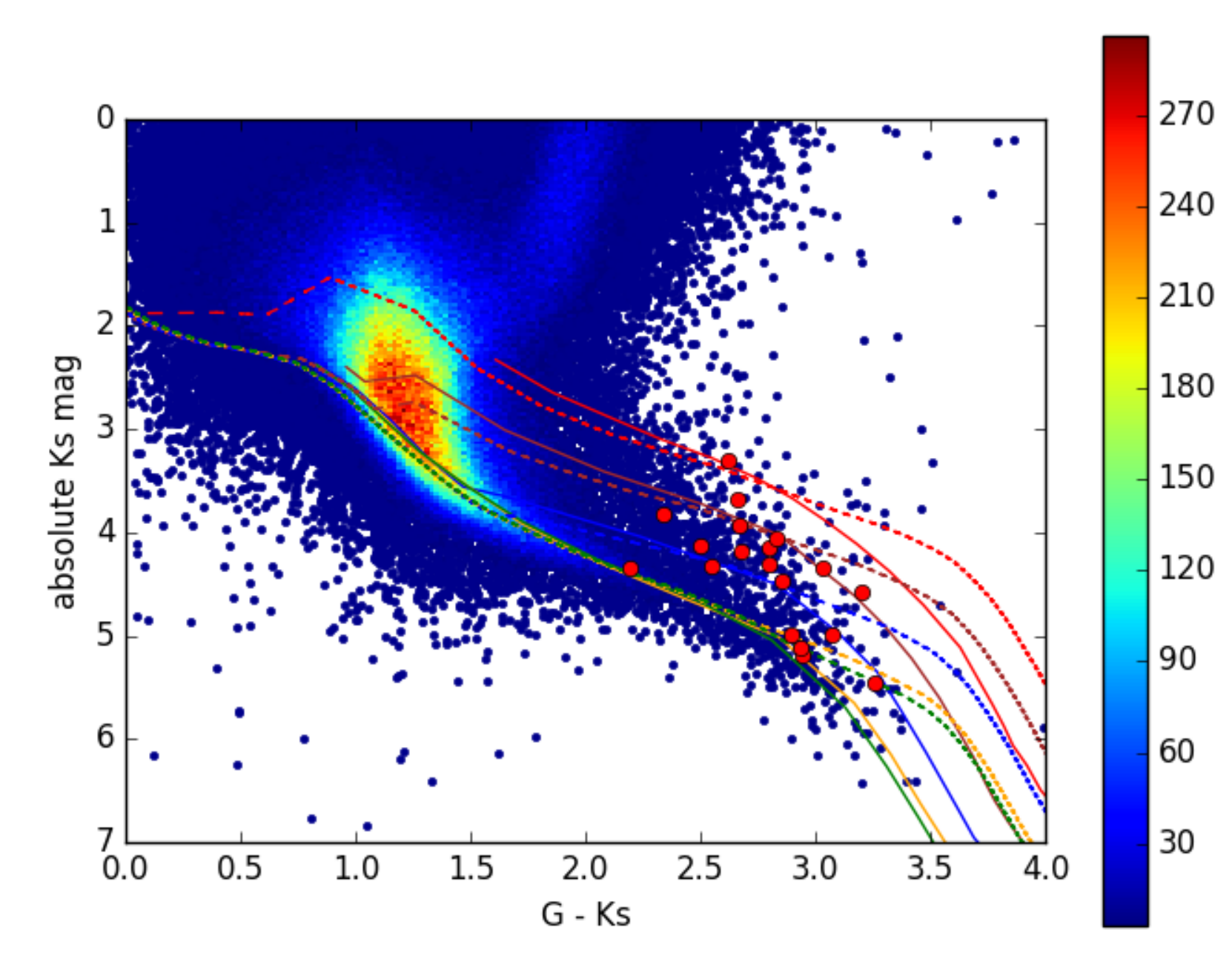}
\includegraphics[height=3.5in,angle=0]{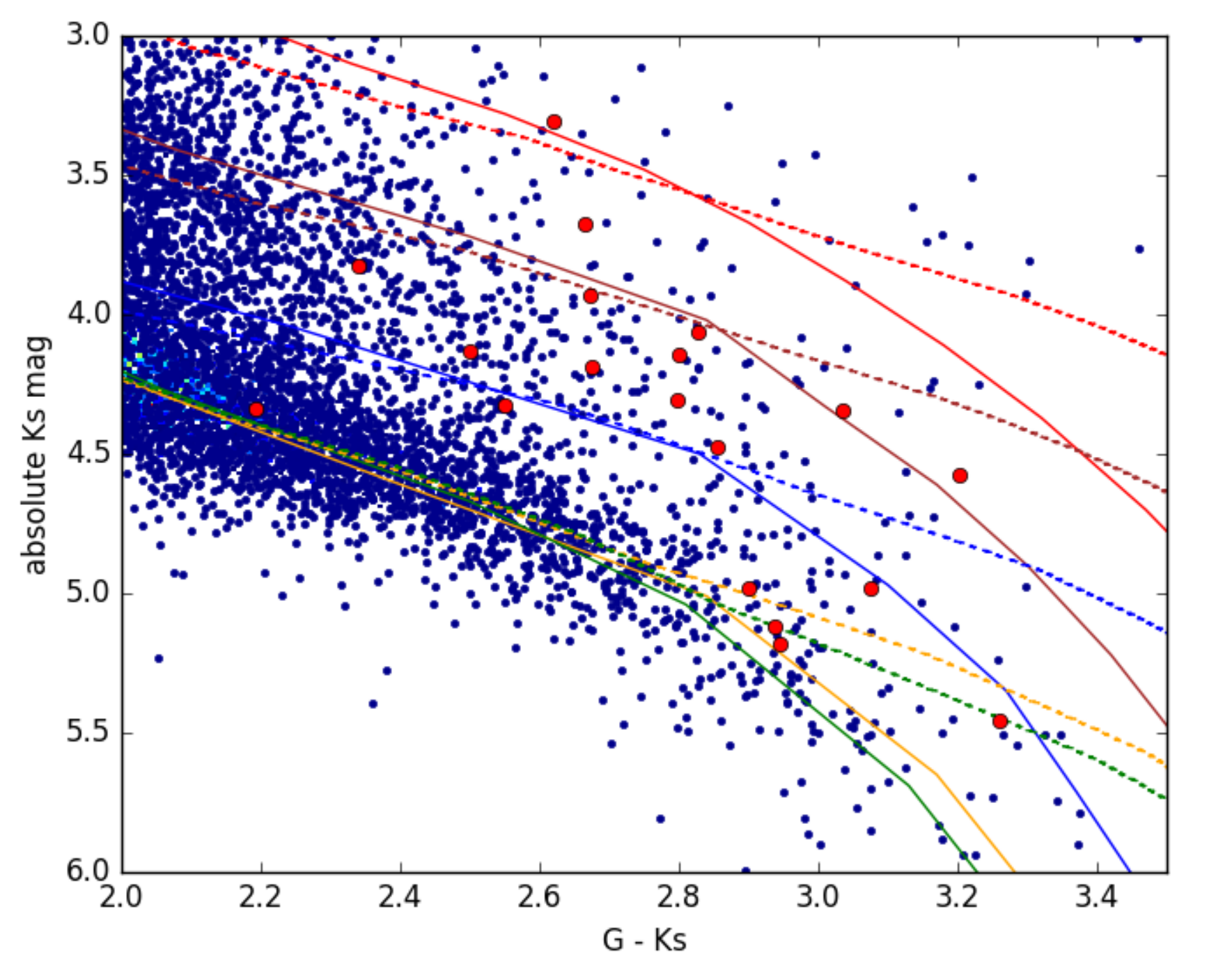}
\vspace{-.18in}
\caption{\footnotesize 
{\it Top panel:} Absolute $K_S$ vs.\ $G-K_S$ CMD for all {\it WISE}- and {\it Galex}-detected objects  in the {\it Gaia} TGAS catalog (rendered as in Fig.~\ref{fig:GvsJmW1}) with $x-$ and $y-$axis ranges as in Fig.~\ref{fig:GaiaNYMGs} and  with the positions of the 19 Table 1 stars indicated as filled red circles.  Isochrones from \citet[][]{Tognelli2011} and \citet[][]{Baraffe2015} are represented as dashed and solid curves, respectively, with colors as in Fig.~\ref{fig:GaiaNYMGs}. {\it Bottom panel:} zoomed-in version of the CMD in the top panel, highlighting the positions of late-type stars. Compare with Fig.~\ref{fig:Good19colorMag}.}
\label{fig:GaiaIsochrones}
\end{figure*}

\begin{figure*}[!h]
\centering
\includegraphics[height=4in,angle=0]{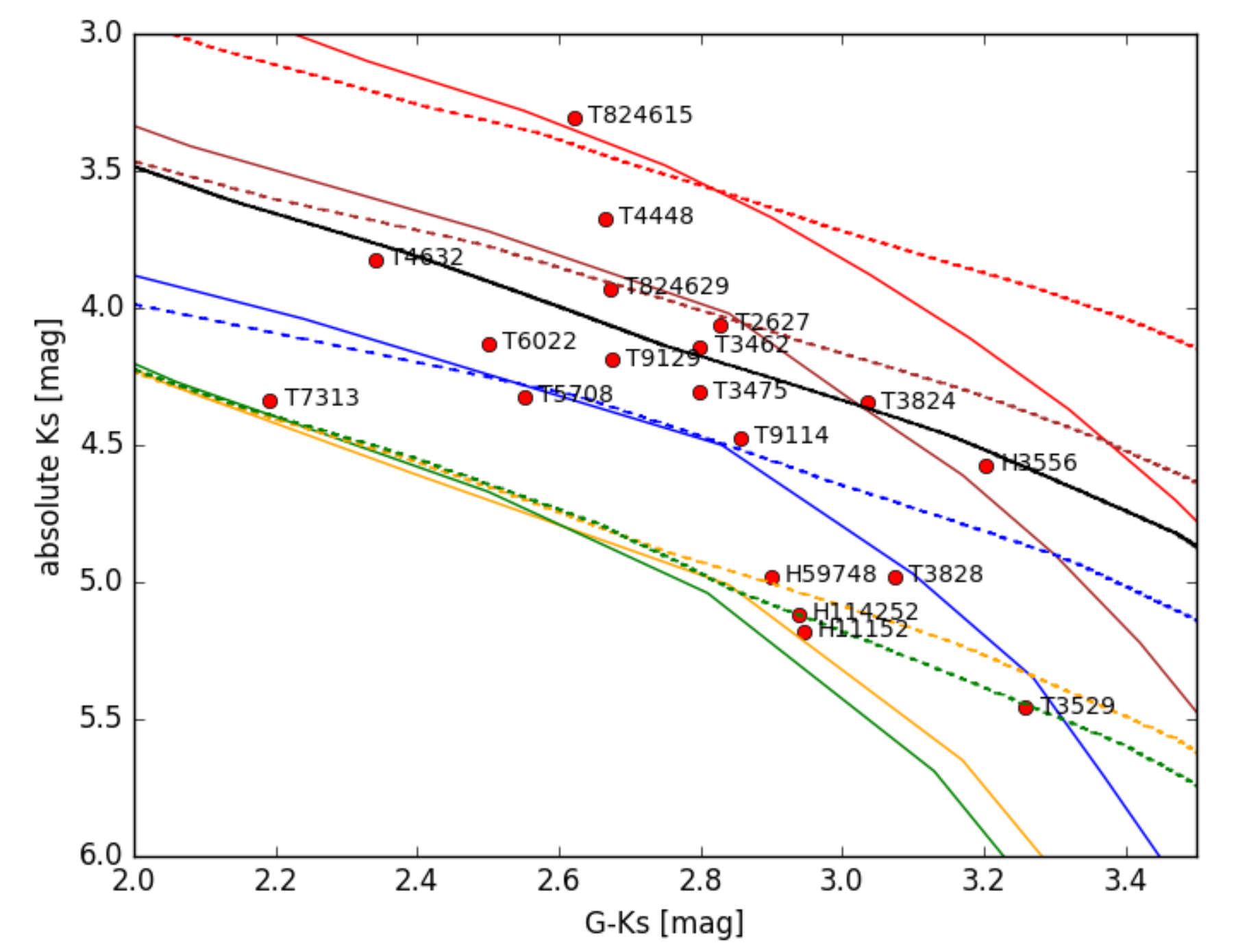}
\vspace{-.18in}
\caption{\footnotesize 
Absolute $K_S$ vs.\ $G-K_S$ CMD for the Table~\ref{tbl:GaiaData} stars (red circles) overlaid with 
isochrones from \citet[][]{Tognelli2011} and \citet[][]{Baraffe2015}, represented as dashed and solid curves, respectively (with color coding as in Figs.~\ref{fig:GaiaNYMGs}, \ref{fig:GaiaIsochrones}). The solid black curve represents the \citet[][]{Tognelli2011} 80 Myr isochrone for unresolved binary systems composed of equal-luminosity components. }
\label{fig:Good19colorMag}
\end{figure*}

\begin{figure*}[!h]
\centering
\includegraphics[height=4in,angle=0]{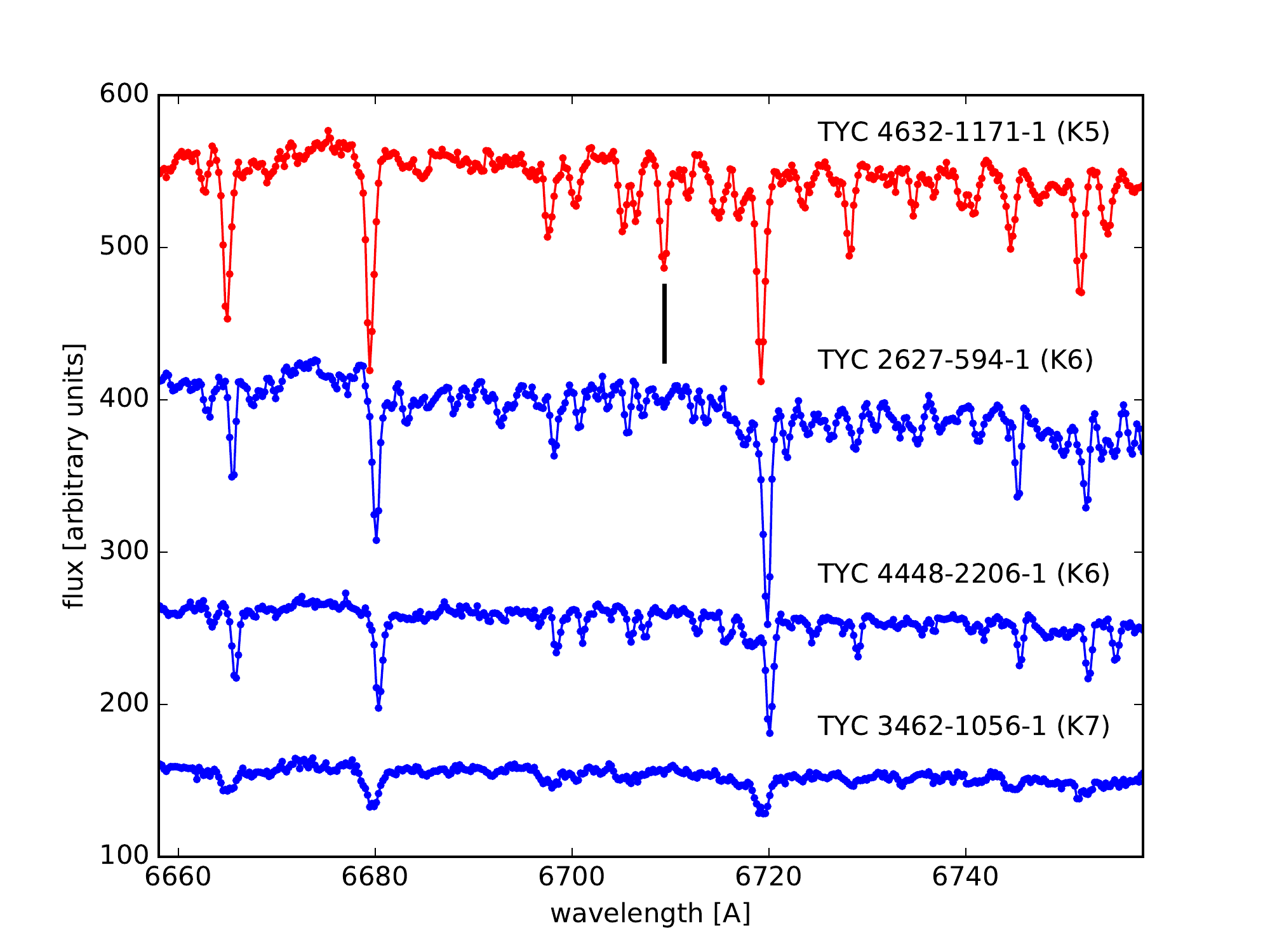}
\vspace{-.18in}
\caption{\footnotesize 
Li {\sc i} $\lambda$6708 line regions of Keck/ESI spectra obtained for four (of five) newly identified young stars found at northern declinations (see \S 4.1.1). Among these high-declination stars, only TYC 4632-1171-1 (top spectrum) displays a clearly detectable Li absorption line.}
\label{fig:sampleSpectra}
\end{figure*}

Comparison of the positions of the 19 Table~\ref{tbl:GaiaData} stars with those of known members of four well-studied NYMGs in a {\it Galex/WISE} color magnitude diagram (\S 3.2; Fig.~\ref{fig:GaiaNYMGs}) indicates that their ages lie in the same general range as these young stellar groups --- i.e., $\sim$8 Myr (TWA) to $\sim$100--150 Myr (AB Dor) --- as expected for stars in the GALNYSS sample. Their positions with respect to pre-main sequence isochrones in a {\it Gaia/2MASS} color-magnitude diagram (Fig.~\ref{fig:Good19colorMag}) yield age estimates for the individual Table~\ref{tbl:GaiaData} stars of from $\sim$10 Myr to $\sim$100 Myr (\S 3.3; Table~\ref{tbl:Summary}). For the subset of (15) stars for which Li {\sc i} absorption line EW measurements are available, the isochrone-based ages are generally consistent with constraints placed on these stars' ages by the Li~{\sc i} EW data. However, a half-dozen stars 
show significant discrepancies between these two (isochronal and Li-based) age determination methods, wherein the Li-based ages (or lower limits on ages) are older than indicated by comparison with isochrones. While such discrepancies would be consistent with these stars' pre-MS status \citep{Jeffries2017}, in at least some cases the differences between isochronal and Li-based ages are likely due to binarity. Several stars lie very close to the main-sequence locus in Fig.~\ref{fig:Good19colorMag}; the pre-main sequence status of these stars (discussed individually in \S \ref{sec:Imposters}) should be regarded with suspicion.

Many of the GALNYSS objects with {\it Gaia/2MASS}-based isochronal ages $\stackrel{<}{\sim}$100 Myr have not previously been considered as candidate young stars in the literature. Eight (of 19) GALNYSS objects that are here newly identified or verified as young stars within $\sim$110 pc are found at declinations north of $+$45$^\circ$, which is highly atypical of members of known groups (\S \ref{sec:NorthernStars}). Indeed, only a handful of the stars identified here as young have $UVW$ space velocities and $XYZ$ positions that appear compatible with membership in known nearby, young moving groups; notable exceptions are HIP 3566 and TYC 9114-1267-1, both of which are likely Tuc-Hor Association members.
Two candidate young stars, HIP 59748 and TYC 3529-1437-1, lie within a mere $\sim$25 pc of Earth (\S \ref{sec:Imposters}). The  isochronal ages of these two stars may be underestimated due to binarity. However, if  confirmed as young, these stars would be particularly attractive targets for direct imaging searches for giant exoplanets. 
 Three stars are members or candidate members of wide binaries (\S\ref{sec:WideBinaries}). These include TYC 9114-1267-1, which is $\sim$830 AU from  Tuc-Hor member V390 Pav, and TYC 8246-2900-1, which is a possible $\sim$47 kAU separation comoving companion to the well-studied, dusty F-type binary HD 113766.

The fact that all (19) GALNYSS stars with parallax distances $D \le 120$ pc in {\it Gaia} DR1 (TGAS) have {\it Gaia}-based isochronal ages in the expected range of $\sim$10--100 Myr, and that the majority of these are evidently {\it bona-fide} pre-main sequence stars, indicates that the GALNYSS UV- and space-motion-based object selection methodology developed by \citet{Rodriguez2011,Rodriguez2013} offers an efficient means to identify young stars near the Sun. These results suggest that hundreds of GALNYSS sample objects are indeed nearby, young stars, and that a significant fraction of these stars likely have not been previously recognized as having ages $\stackrel{<}{\sim}$100 Myr. Once available, the complete {\it Gaia} mission database will provide the means to test these predictions, by confirming or refuting the youth and proximity of the vast majority of the GALNYSS sample stars, as well as to further refine the GALNYSS near/young-star selection strategy and dramatically expand its application. 

\acknowledgments{\it This research was supported by NASA Astrophysics Data Analysis Program (ADAP) grant NNX12AH37G to RIT and UCLA; NASA ADAP grant NNX09AC96G to RIT; and NASA Exoplanets program grant NNX16AB43G to RIT. This work has made use of data from the European Space Agency (ESA)
mission {\it Gaia}\footnote{\url{https://www.cosmos.esa.int/gaia}}, processed by
the {\it Gaia} Data Processing and Analysis Consortium (DPAC \footnote{
\url{https://www.cosmos.esa.int/web/gaia/dpac/consortium}}). Funding
for DPAC has been provided by national institutions, in particular
the institutions participating in the {\it Gaia} Multilateral Agreement. We thank Nathaniel Monson and Siyi Xu for assistance in obtaining and reducing the Keck telescope ESI data, and the anonymous referee for suggestions that improved the paper. J.H.K. wishes to acknowledge the Study Abroad International (SAI) Faculty Fellowship program, and to thank the SAI and Arcetri Observatory staffs, for facilitating the Arcetri residency during which this research was initiated. J.H.K. also acknowledges support from the Merle A. Tuve Senior Fellowship at the Carnegie Institution's Department of Terrestrial Magnetism and a Smithsonian Institution Visitor's Fellowship with the Radio and Geoastronomy Division of the Harvard-Smithsonian Center for Astrophysics. }


\begin{table*}[!htbp]
\tiny
\begin{center}
\caption{\sc {\it Gaia} TGAS Data for GALNYSS Stars with $\pi\ge8$ mas}
\label{tbl:GaiaData}
\vspace{.1in}
\begin{tabular}{lcccccc}
\hline
\hline
Name	&	$\alpha$ & $\delta$ &	Sp.\ type$^a$& $\pi$ & $\mu_\alpha$ & $\mu_\delta$ \\
                & \multicolumn{2}{c}{(J2000)} & & (mas) & mas yr$^{-1}$ & mas yr$^{-1}$ \\
\hline
HIP 3556		&00h45m28.3143s &$-$51d37m34.816s&	M1.5 (1)&	24.57 (0.34) &	99.05 (0.12)&	-58.50 (0.13) \\
HIP 11152		&02h23m26.7534s& +22d44m05.071s&	M1 V (2)&		36.86 (0.34) &	98.51 (0.17)&	-112.54	(0.1)\\
TYC	6022-1079-1&	09h03m33.6062s &$-$20d35m59.067s&	K3 Ve (3)&	21.96	(0.45)&		-61.46	(1.49)&	-3.30	(1.33)\\
TYC	3824-657-1&	11h01m19.1896s &+52d52m23.325s&	K7$^{b}$&	15.69	(0.52)&	-106.75	(0.39)&	-21.53	(1.39)\\
TYC	3828-36-1$^c$&	11h15m53.7148s & +55d19m49.266s&	M0.5 V (4)&	35.03	(0.29)&	-179.28	(0.37)&	-91.51	(0.49)\\
TYC		4632-1171-1&	11h35m34.1586s &+82d59m21.212s& K5$^b$&	14.02	(0.34)&	-59.99	(0.66)&	-25.53	(0.90)\\
HIP 59748$^d$		&12h15m08.4801s & +48d43m56.455s&	M0.5 V (5)&	43.81 (0.85)	& -239.41	(0.07)&	-53.58	(0.08)	\\
TYC		8246-2900-1&	13h06m50.2206s &$-$46d09m56.365s&  K3 Ve (3)&	10.50	(0.41)&		-40.43	(1.87)&	-18.86	(0.51)\\
TYC		8246-1527-1&	13h06m54.3469s &$-$45d41m31.511s&  K5 Ve (3)&	9.11	(0.32)&		-33.24	(1.52)&	-19.02	(0.39)\\
TYC		3462-1056-1&	13h29m32.0979s & +51d42m11.155s&   K7 (6)&	10.56	(0.32)&		-8.87	(0.37)&	-15.58	(0.72)\\
TYC		3475-768-1&	14h18m42.3443s &+47d55m14.986s&	K7 (7)&	12.71	(0.29)&		-25.20	(0.44)&	6.02	(0.70)\\
TYC		7313-1015-1&	15h18m20.4560s &$-$30d56m35.118s&   K5  Ve  (3) &	14.15	(0.32)&		-40.46	(1.13)&	-26.63	(0.47)\\
TYC		3529-1437-1&	18h17m25.0569s &+48d22m03.058s&	M2 (6)&	50.28	(0.88)&		-46.53	(2.97)&	49.81	(3.28)\\
TYC		2627-594-1&	18h21m50.4645s &+32d53m58.807s&	K6$^{b}$&	14.09	(0.38)&		15.68	(1.20)&	-5.06	(1.14)\\
TYC		5708-357-1&	19h15m34.8331s &$-$08d30m20.056s&	K5$^{b}$&	14.98	(0.29)&9.79	(2.18)&	-7.03	(1.91)\\
TYC		4448-2206-1&	19h23m23.2302s &+70d07m38.848s&	K6$^{b}$&	8.94	(0.85)&		33.96	(3.32)&	46.06	(2.05)\\
TYC		9114-1267-1&	21h21m28.9851s &$-$66d55m07.768s&  K7 V  (3)&	31.14	(0.80)&116.27	(1.18)&	-90.87 (1.60)\\
HIP 114252$^e$		&23h08m19.6620s &$-$15d24m36.049s	&M0 Ve  (3) &39.85	(0.25)	&106.94	(0.19)	&-18.99	(0.13)	\\
TYC		9129-1092-1&	23h35m18.8758s & $-$64d33m42.884s  &K6 V (3) &	18.43	(0.71)&	192.83 (1.94)&	-31.07	(1.79)\\
\hline
\end{tabular}
\end{center}

a) References for spectral types: 1) \cite{PecautMamajek2013}; 2)
\cite{Schlieder2012}; 3) \cite{Torres2006}; 4) \cite{Shkolnik2009}; 
5) \cite{Lepine2013}; 
6) \cite{Riaz2006}; 7) \cite{Stephenson1986a}\\
b) Estimated spectral type based on $G-K_S$, 2MASS, and WISE colors.\\
c) GJ 3653\\
d) BD +49 2126\\
e) HK Aqr
\end{table*}

\begin{table*}[!htbp]
\tiny
\begin{center}
\caption{\sc GALNYSS Stars with {\it Gaia} $\pi \ge 8$ mas: Colors, Distances
  and Luminosities}
\label{tbl:DerivedData}
\vspace{.1in}
\begin{tabular}{lcccccccc}
\hline
\hline
Name	&	Sp.\ type & $G$ & $NUV - G$ & $G - K_S$ & $J-W1$ & $D$ &	$\log{L_{\rm bol}}$ & $\log{(L_X/L_{\rm bol})}$ \\
& & (mag) & (mag) & (mag) & (mag) & (pc) & ($L_\odot$) & \\
\hline
HIP 3556	&	M1.5 & 10.83 & 9.12 & 3.20 & 0.97 & 40.70 (0.56) &$-$0.85 &$-$3.55\\
HIP 11152 &		M1	& 10.29 & 7.43 & 2.95 & 0.92 & 27.13 (0.25)	&$-$1.09 &$-$3.85\\
TYC	6022-1079-1 &	K3	& 9.92 & 6.89 & 2.50 & 0.80 & 45.54	(0.93)&$-$0.68 &$-$2.52\\
TYC	3824-657-1&	K7 & 11.40 & 8.51 & 3.04 & 0.98 & 63.73	(2.11)&$-$0.79 & ...$^a$\\
TYC	3828-36-1&	M0.5	& 10.33 & 7.32 & 3.07 & 0.93 &	28.55	(0.24)&$-$1.06 &$-$3.37\\
TYC	4632-1171-1&	K5	&10.43 &	7.08 & 2.23 & 1.09 & 71.33	(1.73) &$-$0.18 &$-$3.47 \\
HIP 59748	&	M0.5	& 9.67 & 9.21 & 2.90 &  0.92 & 22.83	(0.44)	&$-$1.09 & $<$$-$4.4 \\
TYC	8246-2900-1&	K3 & 11.50 & 7.02 & 2.67 & 0.96 & 95.24	(3.72)&$-$0.69 & ...$^a$ \\
TYC	8246-1527-1&	K5	& 11.13 & 7.82 & 2.62 & 0.84 & 109.8	(3.9) &$-$0.46 &$-$3.36\\
TYC	3462-1056-1&	K7	& 11.82 & 7.13 & 2.80 & 0.99 & 94.70	(2.87)&$-$0.64 &$-$3.31\\
TYC	3475-768-1&	K7	& 11.59 & 7.91 & 2.80 & 0.94 & 78.68	(1.80)&$-$0.86 & ...$^a$\\
TYC	7313-1015-1&	K5	& 10.78 & 7.60 & 2.19 & 0.87 & 70.67	(1.60)&$-$0.52 &$-$3.60\\
TYC	3529-1437-1&	M2	& 10.21 & 8.48 & 3.26 & 0.94 & 19.89	(0.35)&$-$1.32 & $-$2.98 \\
TYC	2627-594-1&	K6	& 11.15 & 8.66 & 2.83 & 0.96 & 70.97	(1.91)&$-$0.71 &$-$3.92\\
TYC	5708-357-1&	K5	& 	11.0 & 7.38 & 2.55 & 0.82 & 66.76	(1.29)&$-$0.89 &$-$3.86 \\
TYC	4448-2206-1&	K6	& 11.58 & 8.05 & 2.67 & 0.92 & 111.9	(10.6)&$-$0.40&$-$3.88 \\
TYC	9114-1267-1&	K7	& 9.87 & 9.17 & 2.86 & 0.93 & 32.11 (0.82)	 &$-$0.86 & ...$^b$\\
HIP 114252&		M0	& 10.05 & 7.60 & 2.94 & 0.87 & 25.09	(0.16)	&$-$1.10 & $-2.78$ \\
TYC	9129-1092-1&	K6	& 10.54 & 8.34 & 2.67& 0.87 &	54.26 (2.09)	&$-$0.69 &$-$3.49$^c$\\
\hline
\end{tabular}
\end{center}
NOTES:\\
a) TYC 3824-657-1, TYC 8246-2900-1, and TYC 3475-768-1 lie $\sim$40--50$''$ from the
centroids of low-significance RASS sources. \\ 
b) Secondary of visual binary in which primary likely dominates RASS
X-ray flux. \\
c) Estimated $\log{(L_X/L_{\rm bol})}$ assuming this (primary) star dominates RASS
X-ray flux from visual binary.
\end{table*}

\begin{table*}[!htbp]
\footnotesize
\begin{center}
\caption{\sc Spectroscopic Observations of GALNYSS Stars with {\it Gaia} $\pi\ge8$ mas}
\label{tbl:Obs}
\vspace{.1in}
\begin{tabular}{lccccc}
\hline
\hline
Name & Telescope/Instrument & Date & RV & EW(H$\alpha$)$^a$ & EW(Li)$^b$ \\ 
& & (UT) & km s$^{-1}$ & (\AA ) &  (m\AA ) \\
\hline
HIP 3556$^c$ & MPG/FEROS & 12 Dec.\ 2013 & $+17.1$$\pm$0.7 & ... & $<$30 \\
               & MPG/FEROS & 20 Dec.\ 2013 & $-8.0$$\pm$1.3 & $-0.86$$\pm$0.06 & $<$30 \\
HIP 11152 & IRTF/SpeX & 5 Nov.\ 2013 & ... & ... & ... \\
TYC 3824-657-1 & Lick/Kast & 3 May 2013 & ... & $-0.4$$\pm$0.3 & $<$60\\
TYC 4632-1171-1 & Keck/ESI & 6 March 2017 & $-$17.9$\pm$1.5 & $-0.43$ & 90 \\
HIP 59748 & Keck/ESI & 7 March 2017 & $-$19.4$\pm$1.5 & $-0.50^d$ & $<$40 \\
 TYC 8246-2900-1 & DuPont/BC832 & 21 June 2014 & ... & $-1.41$$\pm$0.18 & 80$\pm$20\\
TYC 3462-1056-1 & Keck/ESI & 7 March 2017 & $-$13.7$\pm$2.2 & $-1.19$ & $<$40 \\
TYC 3475-768-1 & Lick/Kast & 3 May 2013 & ... & $-0.6$$\pm$0.3 & $<$90\\
 &Keck/ESI & 7 March 2017  & $+2.6$$\pm$1.4 & $-1.3$ & $<$40\\
 TYC 3529-1437-1 & Lick/Hamilton & 25 Aug.\ 2012 & $-25.5$$\pm$1.8 & $-2.26$$\pm$0.35 & $<$65 \\
TYC 2627-594-1 & Keck/ESI & 6 March 2017 & $+7.7$$\pm$1.5 & $-0.36^d$ & $<$40 \\
TYC 5708-357-1 & Lick/Hamilton & 27 Aug.\ 2012 & $-27.0$$\pm$2.9 & $-1.31$$\pm$0.05 & 120$\pm$10 \\
TYC 4448-2206-1 & Lick/Hamilton & 26 Aug.\ 2012 & $+20$$\pm$10 & ... & $<$50 \\
                               & Keck/ESI &  6 March 2017 & $+20.3$$\pm$1.4&  0.0 & $<$40 \\
 TYC 9114-1267-1 & DuPont/BC832 & 21 June 2014 & ... & ... & $<$60\\
\hline
\end{tabular}
\end{center}

NOTES:\\
a) Negative EWs indicate line in emission. \\
b) Equivalent width of Li {\sc i} $\lambda$6708 line.\\
c) Double-lined spectroscopic binary (\S 4.2.2); listed RV is for brighter component.\\
d) Self-reversed H$\alpha$ emission, i.e., central absorption
surrounded by emission; listed EW is integrated over emission components only.
\end{table*}

\begin{table*}[!htbp]
\tiny
\begin{center}
\caption{\sc GALNYSS Stars with {\it Gaia} $\pi > 8$ mas: Space Velocities, Galactic Positions, and Ages}
\label{tbl:Summary}
\vspace{.1in}
\begin{tabular}{lccccccccc}
\hline
\hline
Name & sp. type & $D$ & RV$^a$ & $U, V, W$ & $X, Y, Z$ & EW(Li)$^b$ & Age, Li$^c$ & Age, CMD$^d$ & NYMG$^e$ \\
& & (pc) & (km s$^{-1}$) &  (km s$^{-1}$) & (pc) & (m\AA) & (Myr) & (Myr) & \\
\hline
HIP 3556 & M1.5 & 40.7 & $-1.6$ & $-11.2, -18.4, +5.7$ & $9.7, -13.8 ,-37.0$ & $<$30 
& $>$30 & 40--80$^f$ & Tuc-Hor \\
HIP 11152 & M1 & 27.1 & +10.4 &  $-12.5 ,-13.1, -12.3$ & $-19.0, 11.3, -15.7$ & ... 
& ... & 80--100 & \\
TYC 6022-1079-1 & K3 & 45.5 & +20.4 & $-15.7, -18.1, -4.3 $ & $-16.5, -40.3, 13.4$ & 120 
& 30--100 & 35 & Columba?$^g$ \\
TYC 3824-657-1 & K7 & 63.7 & ... & ... & $-31.1, 15.2, 53.5$ & $<$60 
& $>$30 & 20--30 & \\
TYC 3828-36-1 & M0.5 & 28.6 & $-7.6$ & $-15.3, -21.6, -10.0$ & $-13.2, 8.3, 23.9$ & ... 
& ... & 40--60 & \\
TYC 4632-1171-1 & K5 & 71.3 & $-$17.9 & $-8.9, -25.4, -9.1$ & $-34.6, 48.2, 39.6$ & 90 
& 50--100 & 25 & AB Dor?$^g$ \\
HIP 59748 & M0.5 & 22.8 & $-$19.4 & $-14.5, -22.2, -19.4$ & $-6.6, 5.8, 21.1$ & $<$40 
& $>$30 & 80 & \\
TYC 8246-2900-1 & K3 & 95.2 & +17.3 & $-4.3,-26.1, -2.1$ & $53.3, -74.1, 27.2$ & 80 
& $>$40 & 20 & $\beta$PMG?$^g$\\
TYC 8246-1527-1 & K5 & 110 & +9.4 &  $-7.8, -19.8, -5.6$ & $61.3, -85.1, 32.3$ & 470 
& 10--20 & 10 & TWA$^g$ or Sco-Cen? \\
TYC 3462-1056-1 & K7 & 94.7 & $-$13.7 &  $+3.0, -12.8, -8.9$ & $-13.5, 38.6, 85.4$ &$<$40 
& $>$30 & 25 & \\
TYC 3475-768-1 & K7 & 78.7 & +2.6 & $-8.1, -3.6, +4.8$ & 0.3, 35.5, 70.2 & $<$40 
& $>$30 & 35 &  \\
TYC 7313-1015-1 & K5 & 70.7 & $-21.7$ & $-25.1, -6.8, -7.7$ & $60.0, -26.0, 26.7$ & 230 
& 30--50 & 80 & \\
TYC 3529-1437-1 & M2 & 19.9 & $-25.5$ & $-9.5, -23.8, -6.1$ & $4.3, 17.5, 8.5$ & $<$65 
& $>$20 & 40--100 & AB Dor?$^g$ \\
TYC 2627-594-1 & K6 & 71.0 & $+7.7$ & $+4.3, +8.0, -2.6$ & 32.8, 58.0, 24.4 & $<$40 
& $>$30  & 20 & \\
TYC 5708-357-1 & K5 & 66.8 & $-27.0$ &  $-23.8, -13.4, 0.6$ & $58.2, 31.0, -10.7$ & 120 
& 30--100 & 40 & \\
TYC 4448-2206-1 & K6 & 112 & +20.3 & $-32.6, +16.4, -1.4$ & $-20.6, 101.1, 43.3$ & $<$50 
& $>$30 & 15 & \\
TYC 9114-1267-1 & K7 & 32.1 & +3.3 &  $-14.3, -14.4, -10.2$ & $20.8, -13.7, -20.3$ & $<$60 
& $>$30 & 40 & Tuc-Hor$^h$ \\
HIP 114252 & K6 & 25.1 & +2.7 & $-9.3, -5.1, -7.8$ & $ 6.7, 9.1, -22.4$ & ... 
& ... & 80--100 & \\
TYC 9129-1092-1 & K6 & 54.3 & +15.0 & $-35.2, -32.4, -21.5$ & $24.6, -24.0, -42.0$ & ... 
& ... & 30 & \\
\hline
\end{tabular}
\end{center}

NOTES:\\
a) Radial velocities extracted from SIMBAD for stars not
included in Table~\ref{tbl:Obs} as well as for double-lined binary HIP 3556 (see S\ \ref{sec:PrevConsidered}). \\
b) Equivalent widths of Li {\sc i} $\lambda$6708 line. Values obtained
from GALNYSS spectroscopic survey (Table~\ref{tbl:Obs}), except for
TYC 6022-1079-1, TYC 8246-1527-1, and TYC 7313-1015-1, for which
values are from \citet[][]{Torres2006}. \\
c) Age range inferred from plots of EW(Li) vs.\ spectral type presented in \citet[][their Fig.~8]{Murphy2015}.\\
d) Age (or age range) inferred from star's position relative to theoretical pre-MS
isochrones in Fig.~\ref{fig:Good19colorMag}. \\
e) Star is a likely or potential member of this nearby young moving
group. See \S\ \ref{sec:Individuals}.\\
f)  Isochronal age assumes this system is an equal-components binary (\S \ref{sec:PrevConsidered}). \\
g) Age, $UVW$, and $XYZ$ are suggestive of membership, but BANYAN
II calculator \citep{Gagne2014} yields no probability of
membership. See \S\ \ref{sec:Individuals}.\\
h) Based on likely Tuc-Hor membership status of companion (V390 Pav; \S \ref{sec:WideBinaries}).
\end{table*}

\end{document}